%% file: sample-manuscript.tex
\PassOptionsToPackage{dvipsnames,table}{xcolor}

\documentclass[manuscript]{acmart}

\usepackage{booktabs}
\usepackage{textcomp}
\usepackage{graphicx, subcaption}
\usepackage{diagbox}
\usepackage{multirow}
\usepackage{makecell}
\usepackage{float}
\newcolumntype{Y}{>{\centering\arraybackslash}X}
\usepackage{tabularx}
\usepackage{amsmath}
\usepackage{xcolor,colortbl}
\definecolor{maroon}{cmyk}{0,0.87,0.68,0.32}

  \newcommand{\myrowcolour}{\rowcolor[gray]{0.925}}
\newif\ifblackandwhite
\newcommand{\highest}[1]{\color{Maroon}{\textbf{#1}}}%

\DeclareMathOperator*{\argmin}{arg\,min}

\begin{document}






\title{Advancing Location-Invariant and Device-Agnostic Motion Activity Recognition on Wearable Devices}



\author{Rebecca Adaimi}
\email{r_adaimi@apple.com}

\author{Abdelkareem Bedri}
\email{abedri@apple.com}

\author{Jun Gong}
\email{jun_gong@apple.com}

\author{Richard Kang}
\email{runchang_kang@apple.com}

\author{Joanna Arreaza-Taylor}
\email{arreaza@apple.com}

\author{Gerri-Michelle Pascual}
\email{gpascual@apple.com}

\author{Michael Ralph}
\email{iMic@apple.com}

\author{Gierad Laput}
\email{gierad@apple.com}

\affiliation{%
  \institution{Apple}
  \country{USA}
}

\renewcommand{\shortauthors}{Adaimi, et al.}
\newcommand{\jun}[1]{\textcolor{red}{#1}} 

\begin{abstract}
 Wearable sensors have permeated into people's lives, ushering impactful applications in interactive systems and activity recognition. However, practitioners face significant obstacles when dealing with sensing heterogeneities, requiring custom models for different platforms. In this paper, we conduct a comprehensive evaluation of the generalizability of motion models across sensor locations. Our analysis highlights this challenge and identifies key on-body locations for building location-invariant models that can be integrated on any device. For this, we introduce the largest multi-location activity dataset (N=50, 200 cumulative hours), which we make publicly available. We also present deployable on-device motion models reaching 91.41\% frame-level F1-score from a \textit{single} model irrespective of sensor placements. Lastly, we investigate \textit{cross-location data synthesis}, aiming to alleviate the laborious data collection tasks by synthesizing data in one location given data from another. These contributions advance our vision of low-barrier, location-invariant activity recognition systems, catalyzing research in HCI and ubiquitous computing.

\end{abstract}

\begin{teaserfigure}
  \includegraphics[width=\textwidth]{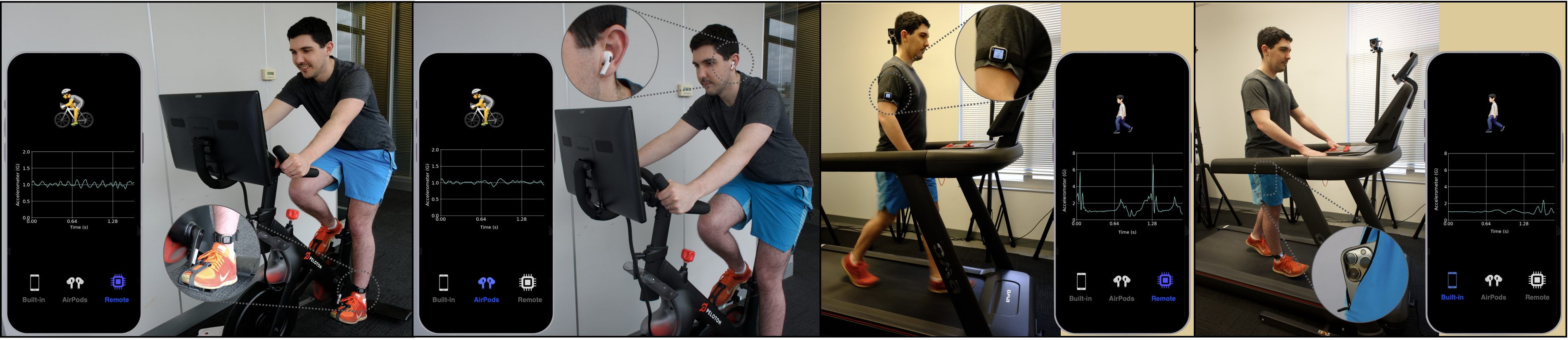}
  \caption{Our motion model is a single on-device model that  enables real-time activity recognition on any device (\textit{e.g.,} watch, phone, and headset) placed anywhere on the body (\textit{e.g.} ankle, head, shoulder, thigh) from a single modality. 
  }
  \label{fig:teaser}
\end{teaserfigure}


\maketitle

\input{main}

\bibliographystyle{ACM-Reference-Format}
\bibliography{bibliography}

\newpage
\appendix
\section{Additional Study Details}
\label{appendix:study_details}

In the study, 50 participants were recruited with 19 identifying as female and 31 as male. Ages ranged between 18 and 60 years. The study took place across two cities in the east and west coast of the United States. Appropriate informed consent was obtained from all participants before participating in the study. 

\subsection{Data Collection Applications}

\begin{figure}[H]
    \centering
        \includegraphics[width=.9\columnwidth]{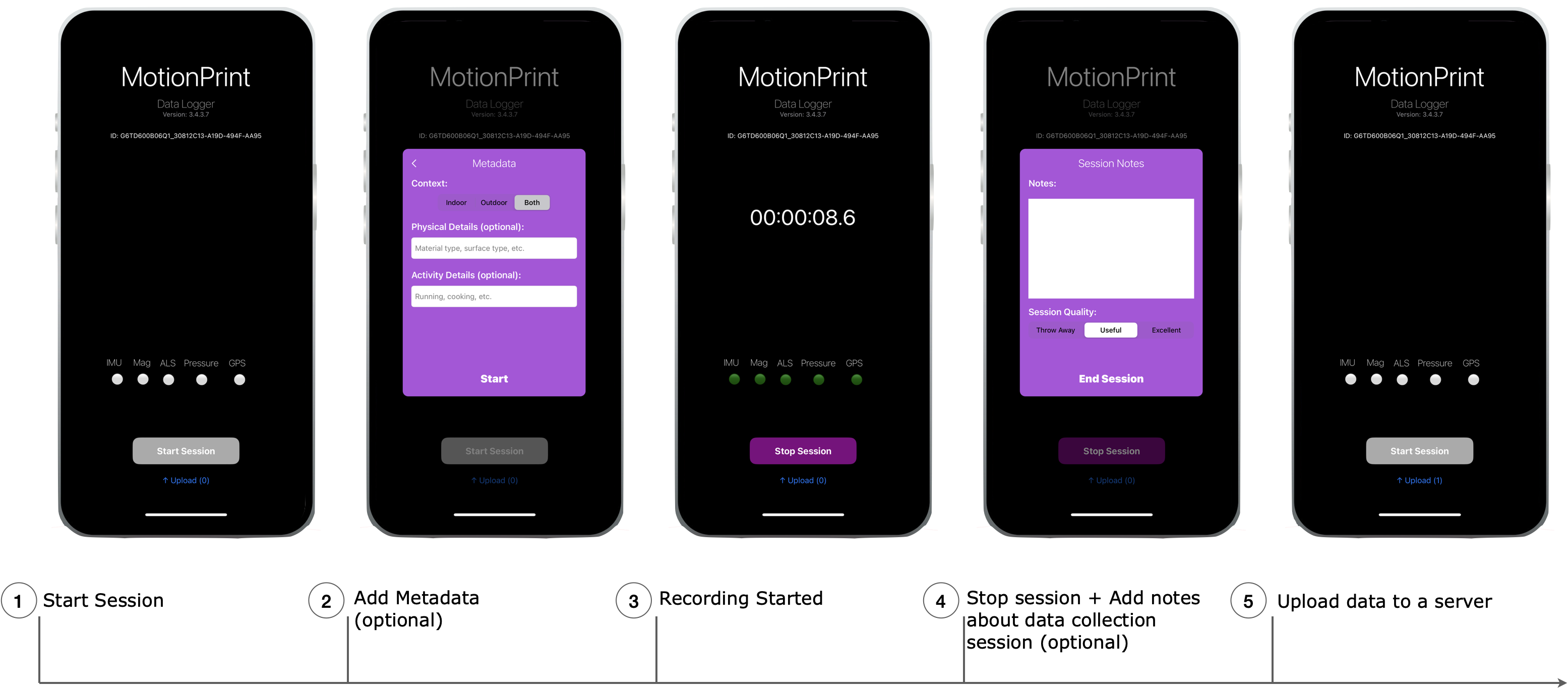}
         \par\bigskip
         \caption{Phone Data Collection Application}
        \label{fig:phone_app}
    \end{figure}

\begin{figure}[H]
    \centering
        \includegraphics[width=.9\columnwidth]{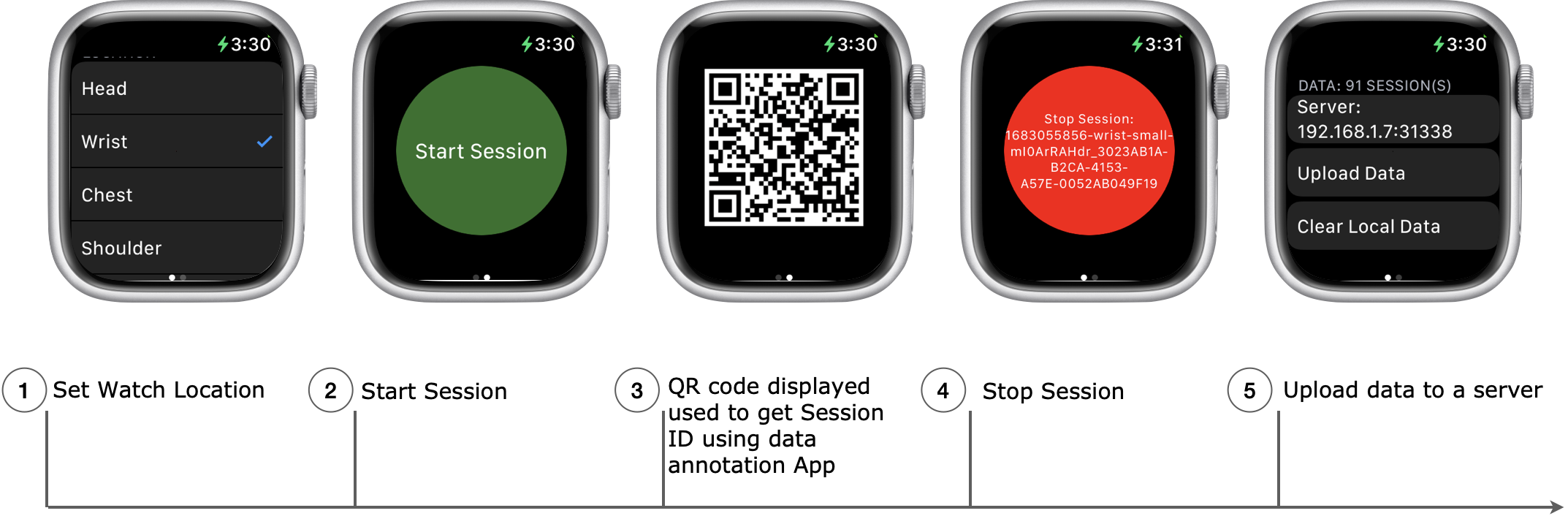}
         \par\bigskip
         \caption{Watch Data Collection Application}
        \label{fig:watch_app}
    \end{figure}

We developed watch- and phone-based iOS-compatible app for data collection. Screenshots of the user interface for both the watch and phone applications can be seen in Figures \ref{fig:phone_app} and \ref{fig:watch_app}. To record the start and stop times of each activity during the session, we also created a separate data annotation app. During the study, a highly-trained proctor attaches the watches to different locations on the participant's body, instructs them to place the phone in their pants pocket, and wear the AirPods in their ears. The proctor ensures that the AirPods are successfully paired with the phone. Before starting the data collection, the proctor initiates logging via the watch app on all watches and the phone app. The data annotation app on a separate device is used to scan the QR code displayed on any of the watches to record the session ID. Throughout the study, the proctor verifies continued operation of all applications. At the end of the session, the proctor concludes the data collection by ending app logging on all devices and uploads the data files from each device to a central server.

\newpage

\subsection{Data Statistics}
\label{sec:app_data_stats}
In any data collection process, it is common to encounter instances of missing data due to various factors. In our study, we experienced cases where data was missing either entirely or partially from certain devices for some participants. For two participants, there was a disruption in the Bluetooth pairing of the AirPods, leading to missing data. Additionally, five other participants experienced missing data from the chest-strapped watch, while another had missing data from the wrist-worn watch. These interruptions were attributed to unexpected crashing of the data collection app. We omit such missing values during analysis as necessary.

\begin{figure}[H]
\centering
\includegraphics[width=.5\columnwidth]{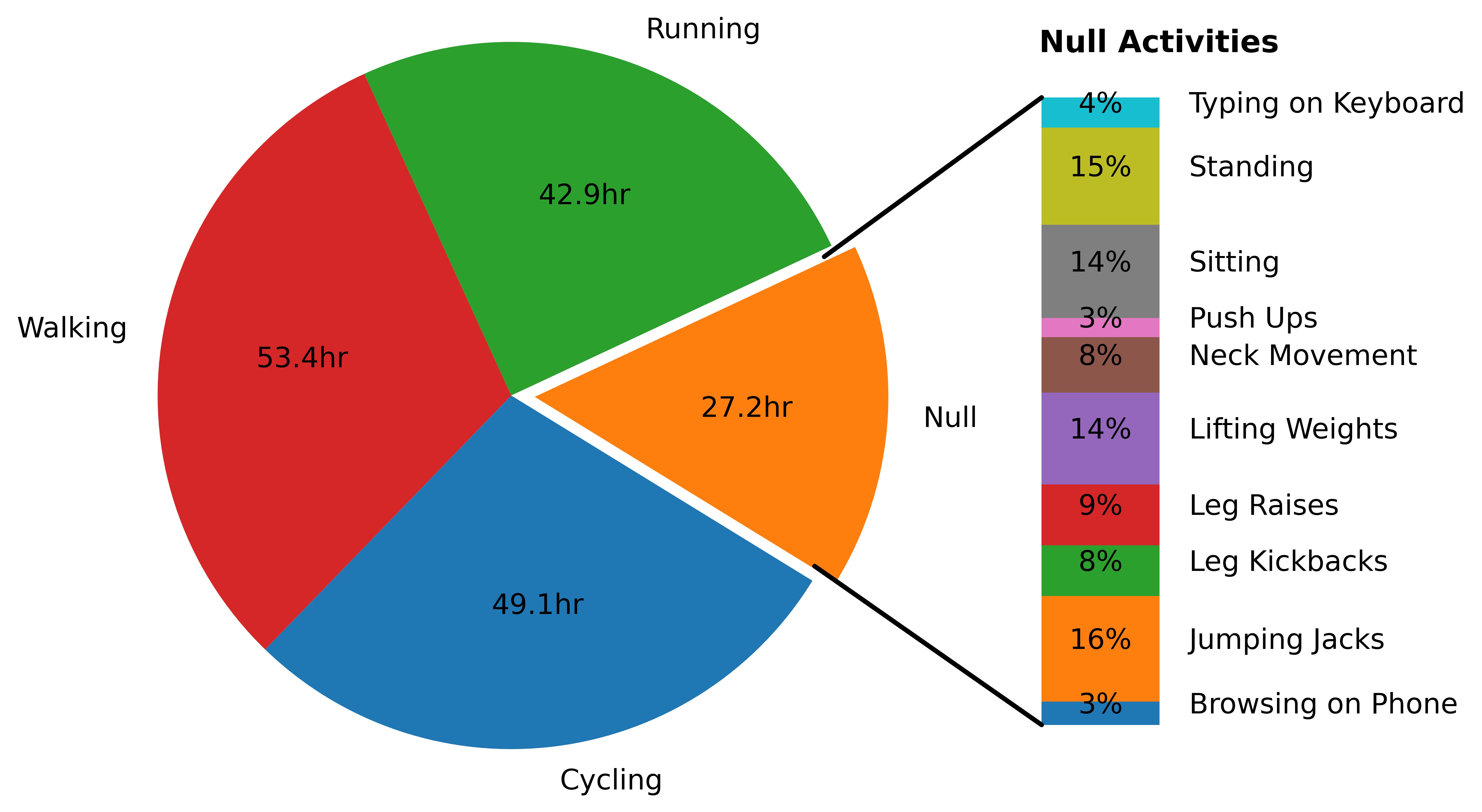}
\caption{Distribution of activities captured in \textit{MotionPrint} dataset.}\label{fig:pie_chart}
\end{figure}
\begin{figure}[H]
    \centering
    \includegraphics[width=1.\columnwidth]{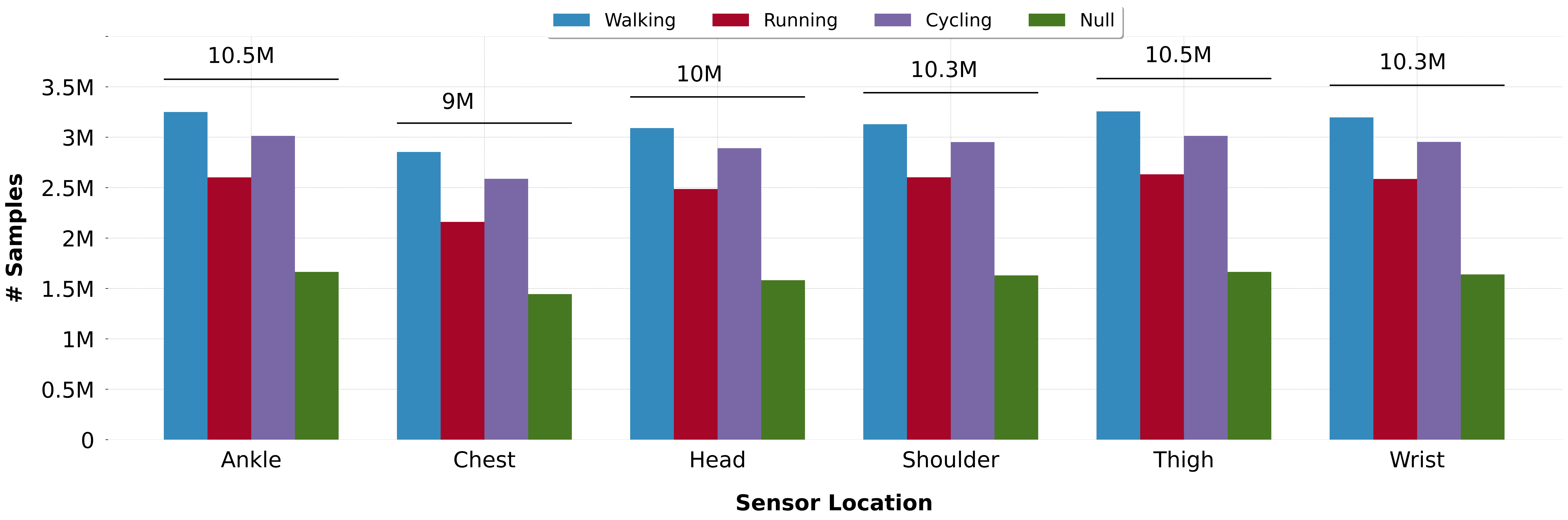}
    \caption{Distribution of the number the of accelerometer data samples captured in \textit{MotionPrint} dataset per location per activity. Note accelerometer data from all locations are resampled to 100Hz.}
    \label{fig:barPlot}
\end{figure}
Upon processing the collected data and excluding instances with missing data, we were able to retain approximately 42–52 hours of recorded motion data for each of the target activities: walking, running, and cycling across all locations combined. This duration encompasses data captured both indoors and outdoors. Furthermore, approximately 27 hours of motion data categorized as "Null" activities were collected. The distribution is visualized in Figure \ref{fig:pie_chart}. Additionally, Figure \ref{fig:barPlot} depicts the number of 3-axis accelerometer data samples collected for each sensor location per activity when resampled to 100Hz. The dataset includes around 10 million samples of accelerometer data per location, with 2.5-3.2 million samples making up each of the target activities.


\newpage
\section{Generalizability Across Varying Sampling Rates}
\label{appendix:SamplingRate}
\begin{table}[H]
    
    \caption{Spectrogram shape for different sampling rates.}
    \centering
    \small
    \begin{tabularx}{\columnwidth}{Y|Y|Y}
    \toprule
        \textbf{Sampling Rate (Hz)} & \textbf{FFT Size} & \textbf{Spectrogram Shape} \\
        \hline
        10 & 26 &  $38 \times 13$ \\ 
        25 & 64 &  $96 \times 32 $ \\ 
        50 & 128 &  $96 \times 64$ \\ 
        75 & 192 &  $96 \times 96$ \\ 
        100 & 256 & $128 \times 128$ \\ 
        \bottomrule
    \end{tabularx}
    \label{tab:samplingRate_parameters}
\end{table}

\newpage
\section{Comparison to Prior Datasets}
\label{append:prior_datasets}

\begin{table*}[!h]
    \centering
    \caption{Summary of multi-device multi-location activity recognition datasets}
        \label{tab:datasets}
        \small
    \begin{tabularx}{\textwidth}{l|l|l|l|X|X}
    \toprule
        \textbf{Datasets} & \textbf{Sensors} &  \textbf{Sampling Rate} &\textbf{\# Users} & \textbf{\# Locations} & \textbf{\# Activities} \\
        \midrule
        Opportunity \cite{chavarriaga2013opportunity} & 3D IMU, orientation & 30 Hz & 4 & 16 \hspace{.25cm} (Left upper arm up, Left upper arm down, Left wrist, Left hand, Right upper arm up, Right upper arm down, Right wrist, Right hand, Right knee up, Right knee down, Right hip, Back, right ankle, left ankle, right shoe, left shoe) &  4 \hspace{.25cm} (standing, walking, sitting, lying) \\
        \midrule
        PAMAP2 \cite{reiss2012creating} & 3D IMU, heart rate & 100 Hz & 9 & 3 \hspace{.25cm}(arm, chest, ankle) &  12 \hspace{.25cm}(standing, walking, sitting, lying, running, cycling, nordic walking, ascending stairs, descending stairs, vacuum cleaning, ironing, rope jumping) \\ 
        \midrule
        RealWorld \cite{sztyler2016body} & 3D IMU, GPS,  sound level & 50 Hz & 15 & 7\hspace{.25cm} (chest, forearm, head, shin, thigh, upper arm, waist) &  8 \hspace{.25cm}(stair up, stair down, jumping, lying, standing, sitting, running, walking)  \\
        \midrule
        DSADS \cite{altun2010human} & 3D IMU & 25 Hz &  8 & 5 \hspace{.25cm}(torso, right arm, left arm, right leg, left leg) & 12 \hspace{.25cm}(sitting ,standing, lying, ascending/descending stairs, walking, running, exercising on stepper, exercising on cross trainer, cycling, rowing, jumping, playing basketball)\\
        \midrule
        MotionPrint (Ours) & 3D IMU & Up to 800 Hz & 50 & 6 \hspace{.25cm}(chest, head, shoulder, wrist, thigh, ankle) & 13\hspace{.25cm} (indoor/outdoor walking, indoor/outdoor running, indoor/outdoor cycling, standing, sitting, typing on keyboard, browsing on phone, lifting weights, push-ups, neck stretching, jumping jacks, leg raises, kickbacks) \\
        \bottomrule
    \end{tabularx} 
\end{table*}    









\end{document}

%% file: main.tex
\section{Introduction}

Sensor-rich mobile and wearable devices are now ubiquitous and more computationally powerful, ushering applications that impact users in many areas, including health and fitness \cite{10.1145/3534582,kheirkhahan2019smartwatch}, safety \cite{10.1145/3410530.3414349}, and accessibility \cite{enabling-hand-gesture}. While numerous advances have been made in the field of Human Activity Recognition (HAR), on-body sensor-based HAR still lags behind other fields, such as sound classification \cite{gemmeke2017audio}, natural language processing \cite{wolf2020transformers}, and computer vision \cite{ho2022cascaded}. 

\begin{figure}[t]
    \centering
    \begin{minipage}[c]{0.49\columnwidth}
    \centering
    \includegraphics[width=\columnwidth]{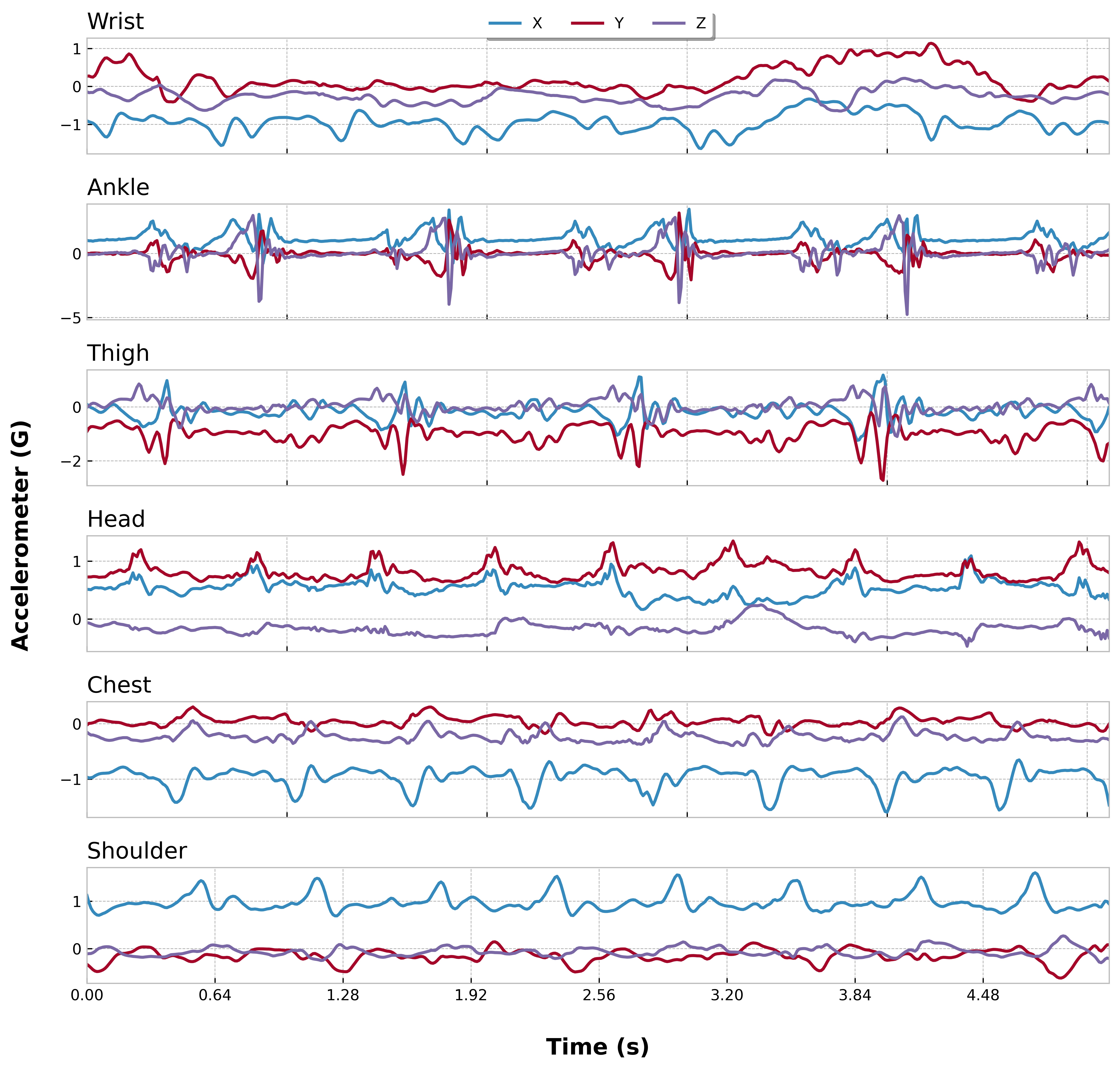}
    \caption{Accelerometer signals of walking, captured from six devices placed at six different on-body locations. Note the similarities \textit{and} the distinct nature of such signals, highlighting the difficulty of training a single model that supports device- and location- diversity.}
    \label{fig:multi_loc_diagram}
    \end{minipage}
    \hfill
    \begin{minipage}[c]{0.49\columnwidth}
    \centering
    \includegraphics[width=\columnwidth]{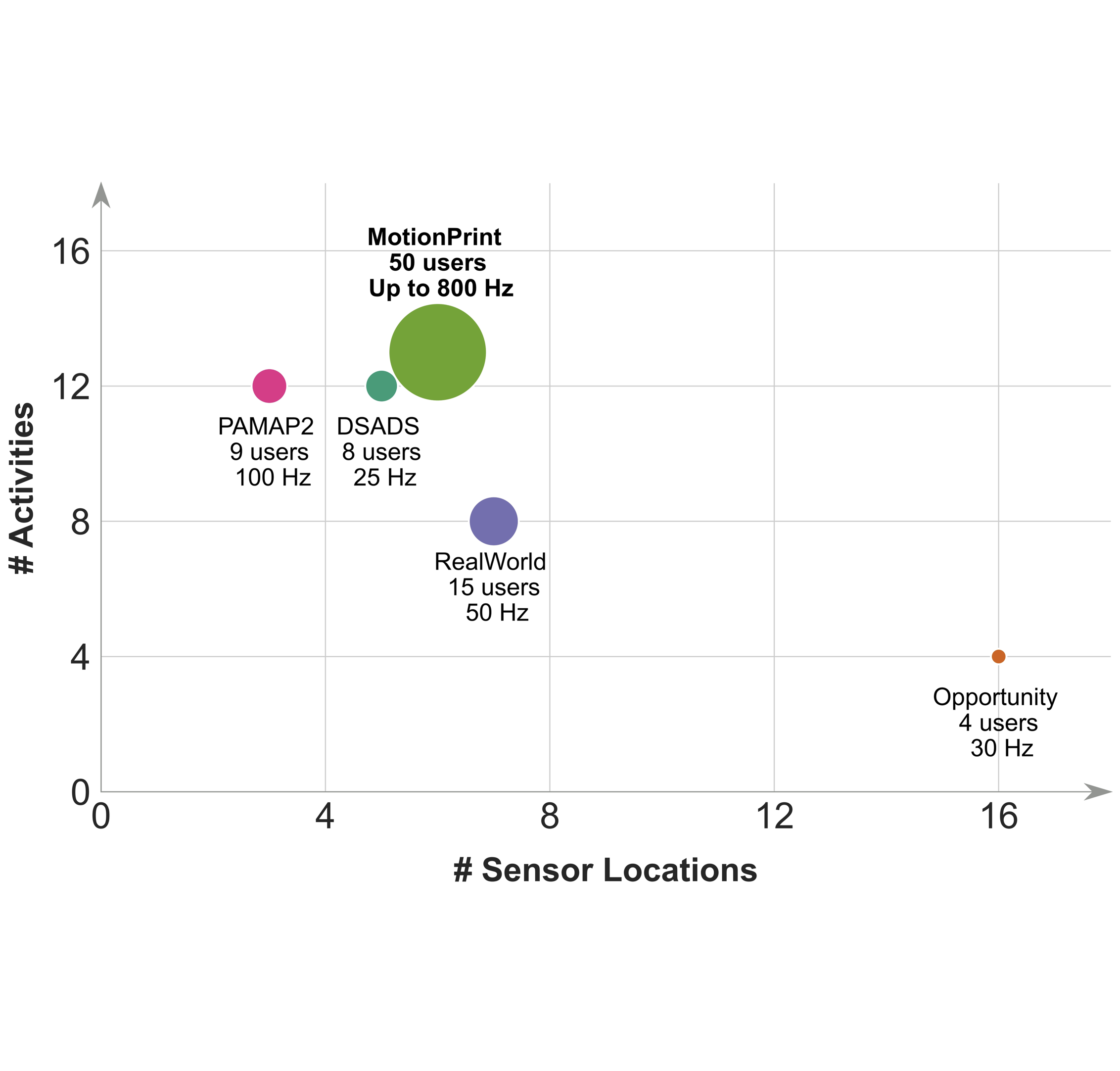}
    \caption{Multi-device, multi-location activity recognition datasets plotted by number of sensor locations (X-axis) and number of activities (Y). Bubble size reflects number of users. Our dataset is the largest multi-device, multi-location sensor-based human activity recognition dataset to date. }
    \label{fig:datasets}
    \end{minipage}
\end{figure}

A major challenge in sensor-based HAR is \textit{sensing heterogeneity}— the ability to generalize to device types, sensor placements, or contextual environments. This places constraints on model generalization, requiring customized models for each location or device platform. For example, in Figure \ref{fig:multi_loc_diagram}, note the significant differences in time-aligned motion patterns captured from six accelerometers placed on different body locations on a walking activity. Model generalization is hard to achieve under such scenarios. Furthermore, sensing heterogeneity necessitates additional data collection each time a new device or placement is introduced, a task that demands significant effort and resources.  In contrast, advanced image recognition models can work on any image captured on most cameras and viewpoints. Likewise, off-the-shelf speech recognition models can work accurately across varying voices and mic configurations. Unfortunately, in the context of sensor-based HAR, such generalization properties are difficult to achieve.



In this paper, we present results from our comprehensive analysis of generalization techniques for motion activity recognition, across different on-body sensor locations and device form-factors.  Achieving this goal is impossible without access to a sizable and diverse multimodal motion dataset. While previous studies have provided sensor-based motion datasets captured from multiple on-body locations, such as  PAMAP2 \cite{reiss2012creating} and Opportunity \cite{chavarriaga2013opportunity}, these datasets are limited in the number of users, sensor locations, and/or activities collected. In response, we introduce the \textit{MotionPrint} dataset, which is (to the best of our knowledge) the largest multi-device, and multi-location human activity dataset collected from inertial sensors (see Figure \ref{fig:datasets} for a comparison). This dataset was collected from 50 people, time-aligned across six locations, from three different types of off-the-shelf devices (\textit{i.e.,} watch, phone, headphones), with sampling rates up to 800 Hz, resulting in 200 cumulative hours of activity data. 

We leverage our \textit{MotionPrint} dataset to perform an in-depth analysis of motion models' capabilities to generalize across a comprehensive range of sensor locations, with an eye towards practicality and deployability. Specifically, we focus on three of the most common full-body activities in the sensor-based HAR literature (\textit{i.e.,} walking, running, and cycling) \cite{ramanujam2021human}, and one of the most common and power-efficient modalities found on wearable devices (\textit{i.e.}, accelerometer). More importantly, full-body activities generally encompass various motion patterns captured across different body parts, each exhibiting distinct intensities corresponding to the specific activity. For instance, cycling entails more pronounced motion patterns in the lower body compared to the upper body. Bearing these characteristics in mind, we carefully chose a selection of the most commonly encountered full-body activities (walking, running, and cycling) for our analysis. The insights learned from this analysis can serve as a foundation for the development of generalized models applicable to other full-body activities.  We also leverage a state-of-the-art on-device architecture, MobileOne \cite{vasu2023mobileone}, and a versatile training scheme to develop a ready-to-use deployable on-device motion model achieving an average of 91.41\% frame-level and 95.17\% activity-level F1-score across all locations. As part of our investigation, we also identify key body locations, which we refer to as \textit{EigenLocations} that offer the best "transfer power" allowing us to develop location-invariant models with a small subset of sensors.

We further investigate the generalization of our \textit{ready-to-use} models to different device form-factors and sampling rates. The demos and video figures accompanying this paper leverage the same model mentioned above (2.1 MB quantized at 16-bit floating point, but can be further optimized to 80 KB on a sparse convolution architecture), which runs on-device and makes predictions on accelerometer data from various devices such as phones, watches, headsets, or even motion data streaming from an off-the-shelf accelerometer deployed on an RPi Zero \footnote{\url{https://www.raspberrypi.com/products/raspberry-pi-zero/}}. We make our \textit{MotionPrint} dataset and \textit{ready-to-use} models accessible to the wider research community, to inspire further research, facilitate benchmarking, and help scaffold applications in ubiquitous computing and interactive systems.

We are also aware of the laborious task of data collection, especially when dealing with multiple sensor placements. We further leverage our large-scale \textit{MotionPrint} dataset to take a step forward in alleviating this challenge. Specifically, we explore the generation of \textit{synthetic data} of inertial sensors from one location given data from another. For example, data collected on the wrist can enable the synthesis of data from other locations (\textit{e.g.,} head, thigh) without the need for additional data collection. To achieve this, we designed a generative Siamese-like AutoEncoder that learns to extract features from each location separately. We take inspiration from Optimal Transport (OT) theory \cite{flamary2016optimal} used for domain adaptation and employ an OT module that learns a transport mapping of features from the source location to a target location. Using this framework, we conduct an exhaustive quantitative and qualitative analysis of our cross-location data synthesis scheme across multiple source-target location pairs. Our results include a discussion of the possibilities it offers and the constraints it imposes.


In summary, our work makes the following contributions:
\begin{itemize}
    \item We present results from a comprehensive investigation on the generalizability of motion models across diverse on-body sensor positions and devices. Subsequently, we identify key \textit{EigenLocations} that facilitate the development of \textit{location-invariant} \textit{ready-to-use} motion activity models.
    \item We present, to the best of our knowledge, the \textbf{largest multi-device multi-location sensor-based human activity dataset}, \textit{MotionPrint}\footnote{\url{www.to-be-added.com}}, strongly labeled, and collected from 50 participants wearing commodity devices placed at multiple on-body locations.
    \item We present results from an in-depth analysis of our \textbf{cross-location data synthesis} technique, where we enable synthetic data generation from one location given data from another.
    \item We make available our on-device motion model that is readily deployable, and illustrate its adaptability across different on-body placements and device form-factors. We hope that this model aids researchers and practitioners with their rapid prototyping and application scaffolding needs.
    
\end{itemize}


\section{Related Work}


In this section, we review several sets of prior work relevant to this paper and highlight what sets apart our work from prior research.

\subsection{Multi-Device and Multi-Location Datasets}
Prior work has contributed to several datasets covering different types of activities captured by various sensors placed at various body positions. In this review, we focus on datasets in the context of everyday movements and activities, such as walking, running, etc. Table \ref{tab:datasets} in Section \ref{append:prior_datasets} in appendix summarizes these existing datasets. Of note, existing datasets have a relatively limited number of users, activities, and/or sensor devices covered, \textit{e.g.} Opportunity \cite{chavarriaga2013opportunity}, DSADS \cite{altun2010human}, and PAMAP2 \cite{reiss2012creating}. The most closely related dataset is the RealWorld dataset \cite{sztyler2016body}, introduced for studying on-body localization of wearable devices. This dataset covers eight locomotion activities captured from 15 participants wearing seven wearable devices at different locations. Our dataset, on the other hand, is a large-scale multi-sensor dataset collected from 50 participants wearing six wearable devices placed at six key body positions where users typically carry a device (chest, head, shoulder, wrist, thigh, ankle). Participants carried out a range of thirteen activities including locomotion activities in both indoor and outdoor environments. 


\subsection{Multi-Device and Multimodal Learning}
Given these prior datasets, we briefly review past research, focusing on work related to multi-device and multi-location configurations for sensor-based HAR. 

\subsubsection{Sensor Fusion}

Sensor fusion has been a widely studied research area for multimodal representation learning \cite{s19173808}. A large number of previous works focused on designing sensor fusion strategies to capture cross-modal interactions and improve classification accuracy \cite{s16010115, 10.1145/3214277, 10.1145/3038912.3052577}. Existing multimodal approaches have evaluated the fusion of modalities at the early, mid, or late stages of model training. These methods have been extensively studied for combining data from a wide range of modalities, such as audio, video, and inertial \cite{10.1145/3534582, 10.1145/3323679.3326513,10.1145/2911996.2912051}. While these strategies have been shown to improve the accuracy of activity recognition, they require the availability of all modalities or devices at inference time. In response, others have investigated cross-modal or cross-view learning approaches, where modality-specific or device-specific models are trained using knowledge distillation or transfer learning from one modality to another\cite{jain2022collossl}.  

\subsubsection{Self-Supervised Learning}
Given the various challenges of collecting labeled human activity datasets, several efforts have been devoted to facilitate knowledge transfer between differing domains in the field of human activity recognition \cite{saeed2019multi}. For instance, Saeed \textit{et al.} \cite{o2015cross} presented a teacher-student network that combines self-training and multi-task self-supervision to learn a robust representation, from the wrist-based unlabeled Fenland dataset, which is then fine-tuned for downstream tasks. Their evaluation included fine-tuning on target datasets that included data collected from the waist, thereby studying performance across sensor locations. Jain \textit{et al.} specifically focused on leveraging unlabeled data collected from multiple devices \cite{10.1145/3517246}. More specifically, leveraging the natural transformation of time-aligned sensor data from multiple devices placed at different locations, they proposed a contrastive learning approach made up of an optimal device selection approach and contrastive sampling algorithm to learn a robust representation on a specific device- or location- anchor. Likewise, Rey \textit{et al.} proposed a method to transfer information from one source sensor location to another target sensor location \cite{fortes2022learning}. Their method requires building a separate encoder for each sensor and applying a contrastive loss to align feature representations. Given the initial success of self-supervised methods for HAR, Haresamudram \textit{et al.} presented a comprehensive assessment of such methods, including studying their robustness to varying sensor locations. However, they only focused on two locations (waist and leg).     
\\\\
Compared to the \textit{pretrain-finetune} paradigm that dominated prior work, we approach this problem with the aim of eliminating the need for any fine-tuning and moving closer towards a \textit{location-invariant}, \textit{device-agnostic} motion model. Thus, we first investigate the challenge of generalizability encountered by motion models when confronted with diverse sensor characteristics, such as sensor placements and sensor types. Then, guided by the results of our analysis, we identify representative sensor locations and build a \textit{ready-to-use} deployable motion model effective on motion data captured from any device irrespective of its location on the body. 

\subsection{Sensor Data Synthesis}

Sensor data collection for HAR is an expensive and laborious task, and has often hampered the field's advancement compared to other domains. Despite numerous efforts in improving human activity dataset collection, the scale of typical datasets remains small \cite{chavarriaga2013opportunity,reiss2012creating}. Apart from data augmentation methods \cite{fawaz2018data}, several data generation solutions have emerged leveraging the success of generative adversarial networks (GANs) in image synthesis \cite{goodfellow2020generative}. SenseGen \cite{inproceedings}, SensoryGAN \cite{wang2018sensorygans}, and WGAN \cite{alharbi2020synthetic} are such examples of models used for generating synthetic sensor data from noise input. Other efforts have explored the generation of virtual sensor data from other rich sources, such as video \cite{huang2018deep,kwon2020imutube,rey2019let} and motion capture \cite{xiao2021deep}. 

In a similar vein, we extend this line of research, albeit in a different context. Based on the observation that multiple devices on a user's body capture the same physical activity from different perspectives, we investigate the feasibility of leveraging data from one location as a supervisory signal for data synthesis for another location. As such, we propose a cross-location data synthesis approach that leverages data from one location as a supervisory signal for data synthesis from another location. We leverage \textit{optimal transport}, a mathematical tool used for comparing distributions and finding an optimal transport plan for matching distributions \cite{flamary2016optimal}. This method, formulated as an entropy regularized problem and efficiently solved by the Sinkhorn-Knopp algorithm \cite{cuturi2013sinkhorn}, has been effective in computer vision in solving various machine learning problems, such as domain adaptation \cite{flamary2016optimal,redko2019optimal}, dimensionality reduction \cite{gautheron2019feature}, and generative models \cite{park2018representing}. However, in sensor-based HAR, optimal transport has been used once to learn a mapping of substructures of clustered activities from a source dataset to a target dataset, effectively applying cross-domain adaptation \cite{lu2021cross}. Considering the novelty inherent in the research problem of \textit{cross-location data synthesis}, our primary objective in this study was to explore its viability and acquire insights into both the potential advantages it offers and the limitations it entails.

\begin{figure}[t]
   \centering
    \includegraphics[width=.6\columnwidth]{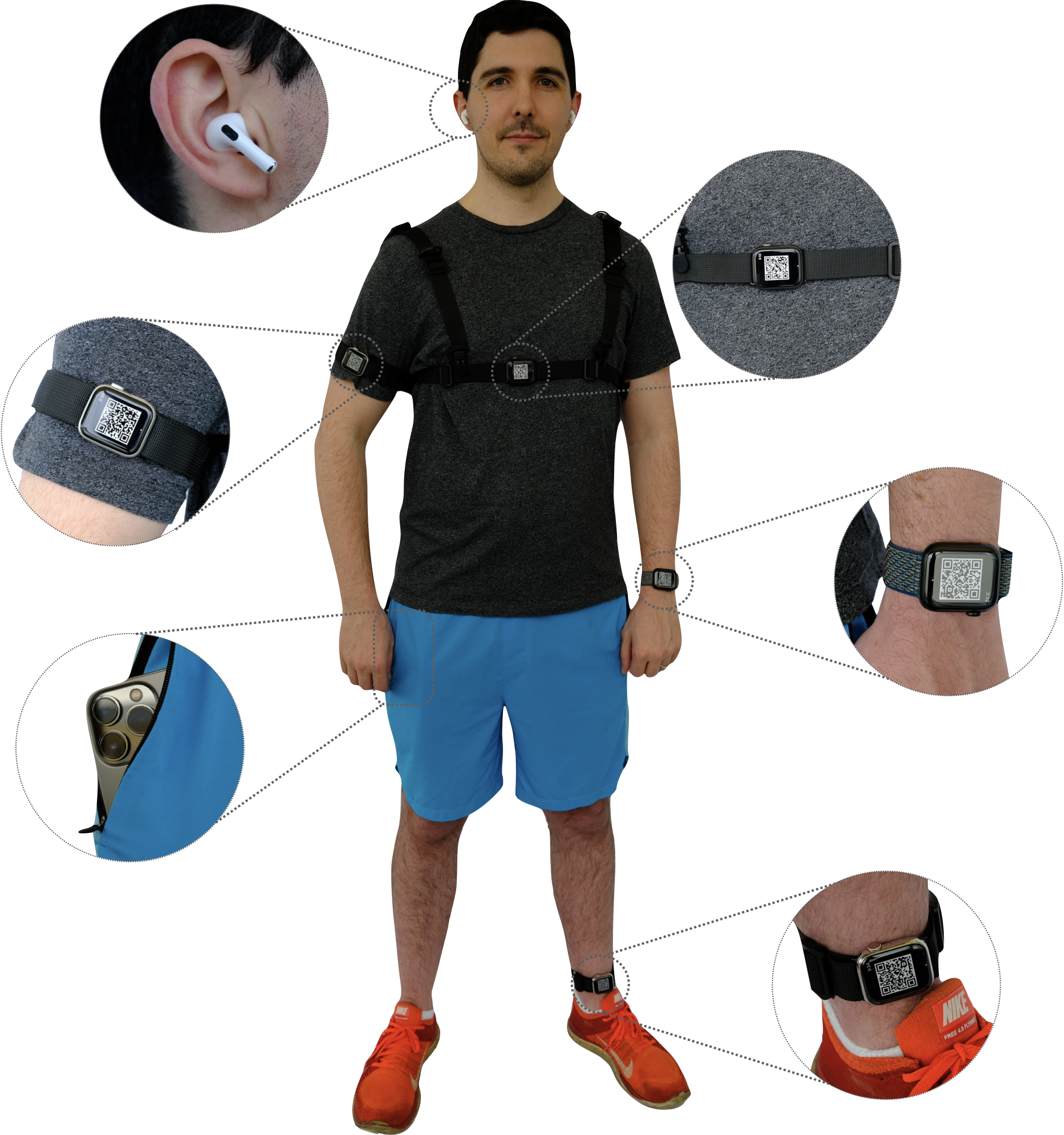}
    \caption{ Our \textit{MotionPrint} data collection apparatus. Participants wore four watches, earbuds, and a phone in their pocket, collecting data from six on-body locations (wrist, shoulder, head, chest, thigh, and ankle).}
    \label{fig:study_diagram}

\end{figure}

\section{MotionPrint Dataset}
\label{sec:study}

In this section, we describe in detail our large-scale, strongly-labeled, multi-device, multi-location motion dataset, which we reference as the \textit{MotionPrint} dataset. With this dataset, we sought to capture inertial sensor data from different wearable devices placed on distinct on-body locations.

\subsection{Apparatus}
\label{sec:apparatus}
In order to capture motion data on a large scale, we employed a comprehensive data collection setup comprising various devices, including an Apple iPhone 13, which was placed inside the pocket of each participant and four watches (Apple Watch Series 7 and 8, aluminum and stainless steel). These watches were strategically fastened onto the wrist, shoulder, chest, and ankle areas of the participants. Additionally, participants wore a pair of Apple AirPods Pro in-ear. Custom modifiable straps were designed to strap the watches to the different on-body locations. Figure \ref{fig:study_diagram} shows the complete data collection setup. Accelerometer data was collected from all devices, with the sampling rate set to their maximum configurations: watches were configured to 800 Hz, and the phone and AirPods set to 100 Hz with a sensitivity of $\pm 16g$. 

Dedicated iOS applications were developed for both the watches and the phone. These applications were designed to capture inertial data from the built-in sensors of each device and store them locally. Moreover, data collection by the AirPods Pro was integrated into the same phone application. After pairing the Airpods Pro to the phone via Bluetooth, the inertial data was streamed to the phone during data collection. Figures \ref{fig:phone_app} and \ref{fig:watch_app} in the appendix present screenshots showcasing the user interface of the watch and phone applications. This interface served as the primary means for data collection and management. 

To annotate the data, a separate data labeling application was designed and installed on a separate phone device. The key features of the app are recording the start and stop times of each activity using the global Unix timestamps \textit{Time-Interval-Since-1970}. This not only helped segment the activities accurately, but also helped synchronize the sensor data from all devices. This dedicated tool streamlined the process of recording the activities and ensured consistency in labeling across all data collection sessions.

\subsection{Protocol}
The dataset was designed under a semi-natural data collection protocol conducted across two cities in the east and west coast of the United States. 50 participants were recruited, consisting of 19 female and 31 male individuals. The study duration ranged approximately from 60 to 75 minutes. Participants were instructed to wear the six devices discussed in Section \ref{sec:apparatus} following the set of predefined on-body locations, as shown in Figure \ref{fig:study_diagram}. In order to increase variability, we do not specify which side (left or right) to wear or carry the watches and phone. However, in order to capture full-body motion, for each of the wrist-shoulder and ankle-thigh pairs, we asked participants to wear them on opposite sides. For instance, if a participant wore a watch on their right wrist, they were instructed to strap the other device on their left shoulder, and vice versa. 

During the study, participants were asked to engage in a series of three common activities: walking, running, and cycling. Each activity was performed for a duration of 5 minutes, both indoors and outdoors, in order to capture variations in motion patterns. For indoor activities, a treadmill and a stationary bike were used (Figure \ref{fig:teaser}). In addition to the three target activities, participants were also instructed to perform a set of five miscellaneous non-target activities, one minute each, forming a "null" class. These five activities were selected from a predefined set of 12 activities. The selection was made by randomly picking one activity from each of the following 5 categories: (1) idle activities (standing or sitting), (2) wrist-related activities (typing on a keyboard or browsing on a phone), (3) upper-body activities (lifting weights or push-ups), (4) lower-body activities (kickbacks or leg raises), and (5) full-body activities (jumping jacks). By incorporating a variety of activities and capturing data from both indoor and outdoor settings, the \textit{MotionPrint} dataset aims to provide a large-scale representation of motion patterns, with a focus on the most common daily activities. 

Using the applications discussed in Section \ref{sec:apparatus}, data was captured continuously throughout a session, with a highly-trained proctor monitoring the session and recording each activity's start and stop times using the data annotation app. Overall, we collected over 30 hours of data per device (8-9 hours of walking, running, and cycling each), for a cumulative total of 200 hours and 408 million accelerometer samples across all devices and all users. Additional information regarding data statistics can be found in Section \ref{sec:app_data_stats} of the appendix.

\subsection{Additional Platforms and Sensor Locations}
\label{sec:unseen_loc_platforms}

In order to further assess the generalizability of a motion model on unseen sensor locations and other sensing platforms, we conduct additional data collection from unseen sensor locations as well as different sensing platforms. Similar to the \textit{MotionPrint} data collection, we recruited 5 participants (outside of the 50 participants) and equipped them with watches placed at three new on-body locations: Hat, Belt, and Shoe. The specific attachment points were the bill of the cap, the belt buckle, and the outward-facing shoe collar, respectively. The watch was mounted onto a clamp that can be attached to any on-body location, with data being sampled at 800 Hz. At the same time, an off-the-shelf sensor board, an RPi Zero equipped with a ISM330DHCX IMU, representing a different sensing platform than what the model is trained on, is used. The sensor specifications, sampling rate (100 Hz) and sensitivity ($\pm 16g$), were tuned to closely match the sensor configurations used in our \textit{MotionPrint} dataset. For data collection, we attached an RPi device to the Wrist and Ankle using a custom band. For this round of data collection, participants performed the target activities (walking, running, and cycling) indoors for 5 min each as well as a mix of five one-minute "other" activities. This resulted in around 8 hours of motion data, or 15.6 million total samples of accelerometer data across all participants and all unseen locations and sensing platforms.


\section{Ready-to-use Motion Activity Model}
\label{sec:universal_model}
As mentioned in the introduction, one of our main objectives is to lower the \textit{floor} for adoption of HAR-based motion models. In response, we analyze the generalizability of motion models in an effort to deepen our understanding of model generalizability across sensor locations, devices, and sampling rate. As a result of this analysis, we offer insights and answers to the \textit{HAR Universality Hypothesis}, building a \textit{readily-deployable} motion model that can be immediately deployed on-device, works with any accelerometer, at any on-body location. In this section, we discuss the modeling components and results of our analysis.   

\subsection{Motion-Based Spectrogram Images}
\label{sec:Spec_feature_set}
Unlike images, raw time-series data by itself is difficult to visually interpret. Likewise, when analyzing time-series data, important information is manifested in both time and frequency domains. Thus, to represent such signals in both time and frequency domains, we used Short-Time-Fourier-Transforms (STFT) to convert accelerometer signals into spectrograms, which can be viewed as image representations of raw signals. Moreover, we eliminate sensitivities to sensor orientation by computing the magnitude of the accelerometer data before extracting spectrograms. The eventual spectrograms as a result of this process are what we call \textit{spectrogram image feature sets}, which is equivalent to a 1-channel image representation in the vision domain.

To create our spectrogram image feature sets, we first normalize the data per-device and per-channel (per-axis) to zero mean and unit variance and compute its magnitude acceleration. We then process each data sequence with a sliding window, using a  size of 5.12 seconds and a step size of 0.64 seconds. For each frame, we extract a spectrogram image using an FFT length of 256, and a 128-length window with 125 overlap. This results in a $128 \times 128 \times 1$ spectrogram image from each sensor device or location. The spectrograms in turn are normalized using a 99th percentile scaling, while clipping all values to be within [0,1]. 

\begin{figure*}[t]

    \includegraphics[width=\columnwidth]{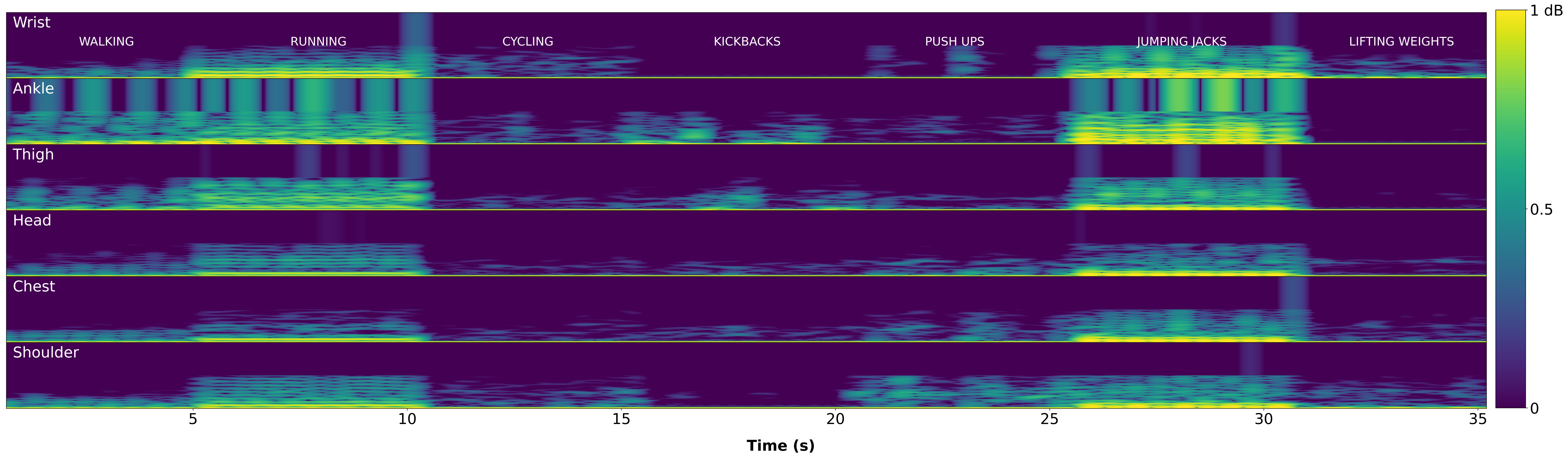}
    \caption{Spectrogram of acceleration signals from six on-body sensor locations for each of walking, running, cycling, and four "other" activities. We only visualize a subset of the "other" activities for clarity. Note the distinctive frequency information captured for each activity across all locations.}
    \label{fig:overview}

\end{figure*}

The spectrogram image features from different locations represent varying views of the same activity. Contrary to prior multi-device HAR methods where an early or late fusion approach is applied, we treat data from each device as a separate input sample. This multi-source data is equivalent to having augmented views of the same activity. As such, the model learns a highly-generic representation of motion activity without any knowledge of device-type or sensor location. Figure \ref{fig:overview} shows a spectrogram snapshot of acceleration data captured from  various locations and devices during each activity. Finally, we use these post-raw spectrogram image feature sets to train and build the motion models for our analysis.

\begin{figure*}[t]

    \includegraphics[width=\textwidth]{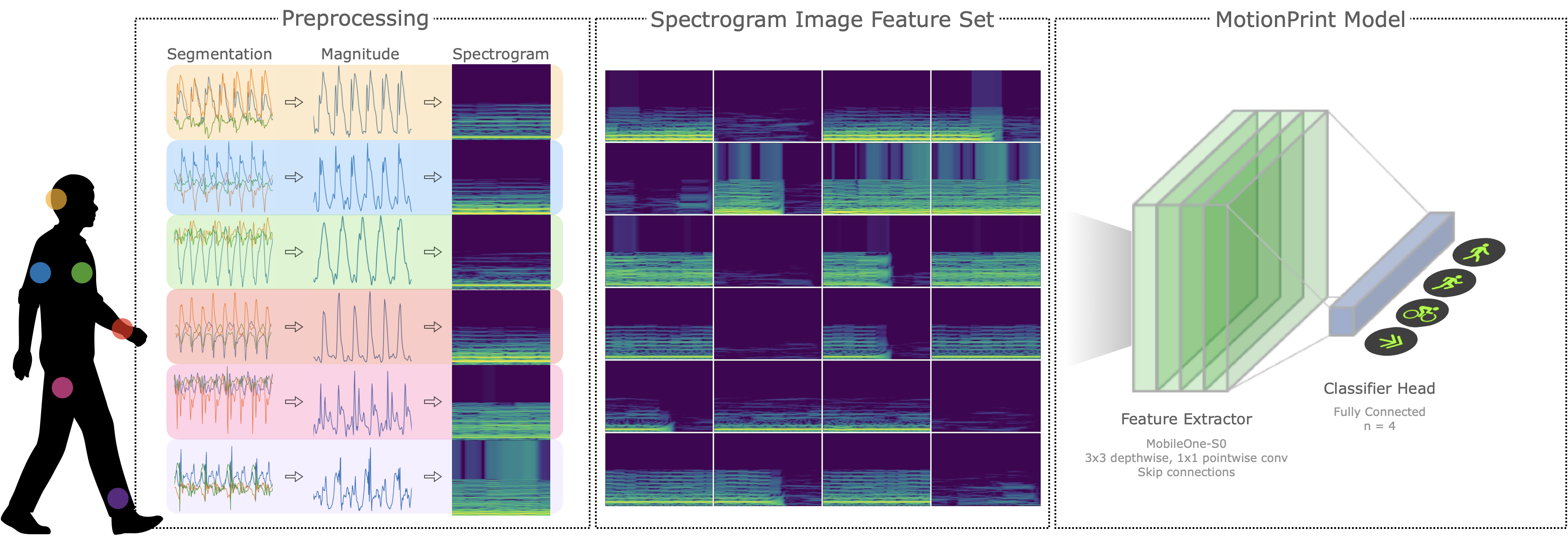}
    \caption{Ready-to-use motion model overview. Leveraging our \textit{MotionPrint} dataset, we create spectrogram image feature sets as a flexible representation for training our motion models. We leverage the MobileOne\cite{vasu2022improved} architecture as our base feature extractor, providing state-of-the art accuracy and efficiency gains for on-device deployment.}
    \label{fig:network_architecture}

\end{figure*}

\subsection{Architecture}


To learn from our spectrogram image feature sets, we modified the MobileOne \cite{vasu2022improved} architecture by using spectrogram images as input for recognizing motion activity. It leverages state-of-the-art reparameterization for efficient inference, and is optimized for on-device compute. CNN-variants have been widely used for visual datasets (i.e., width × height × color channel), and in our case, we represent our motion activities as image-like spatio-temporal patches of accelerometer data. Our architecture consists of two-stage MobileOne blocks, with a kernel size of $3 \times 3$ and $2 \times 2$ stride. A ReLU activation, batch normalization, and skip-connection is applied between every convolutional layer. Finally, the output of the last block is flattened and passed to a classifier consisting of three linear fully-connected layers of 512, 512, and 4 units respectively. The last layer outputs a score for each of the four classes (walking, running, cycling, and other). A dropout layer ($p=0.5$) is added before the last fully-connected layer to avoid overfitting. An illustration of our network architecture is offered in Figure \ref{fig:network_architecture}. 

\subsection{Training}
\label{sec:universal_train_info}
We randomly split our spectrogram feature sets into training (35 users), testing (10 users), and validation (five users) to ensure cross-user evaluation. Following the standard training of deep learning models, we train our model to minimize cross-entropy loss. We trained our model for 500 epochs using an Adam optimizer, with a $1e-5$ learning rate. The model with the best results on our validation set is saved and evaluated on our testing set. 


\begin{figure*}[t]
    \centering
    \centering
    \includegraphics[width=\textwidth]{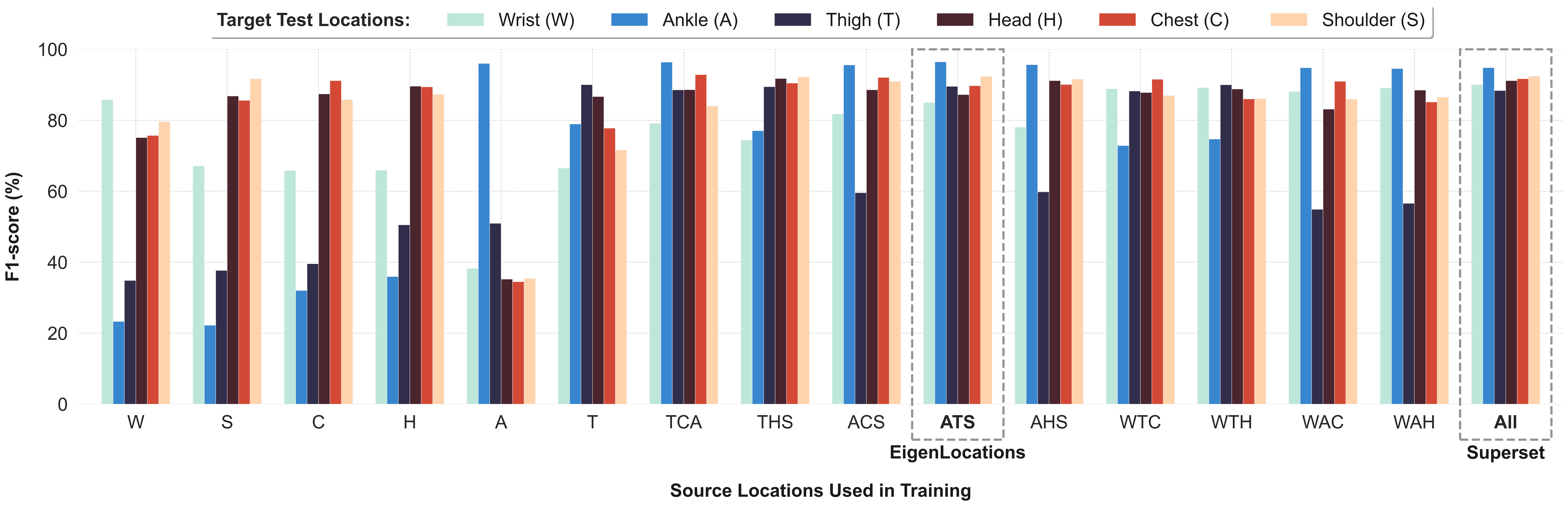}
    \caption{Recognition performance comparison of our motion model trained on data from one or more locations. When trained on all locations, the model achieves high classification performance across all locations. Our exhaustive analysis shows that Ankle-Thigh-Shoulder locations serve as crucial \textit{EigenLocations}, facilitating the creation of a \textit{location-invariant} model that  achieves comparable performance against a  model trained on the superset of all sensor locations.}
    \label{fig:cross_location}
\end{figure*}

\subsection{Generalizability Analysis}

Our next goal was to assess the generalizability property of our motion model across two potential sensing heterogeneities—  variations in on-body location and sampling rate. From this, we share insights and propose improvements to further optimize the plug-and-play ethos and universality property of our model. For all evaluations, we compute the macro F1-score averaged over our target classes. 

\begin{figure}[t]
    \centering
    \includegraphics[width=.8\columnwidth]{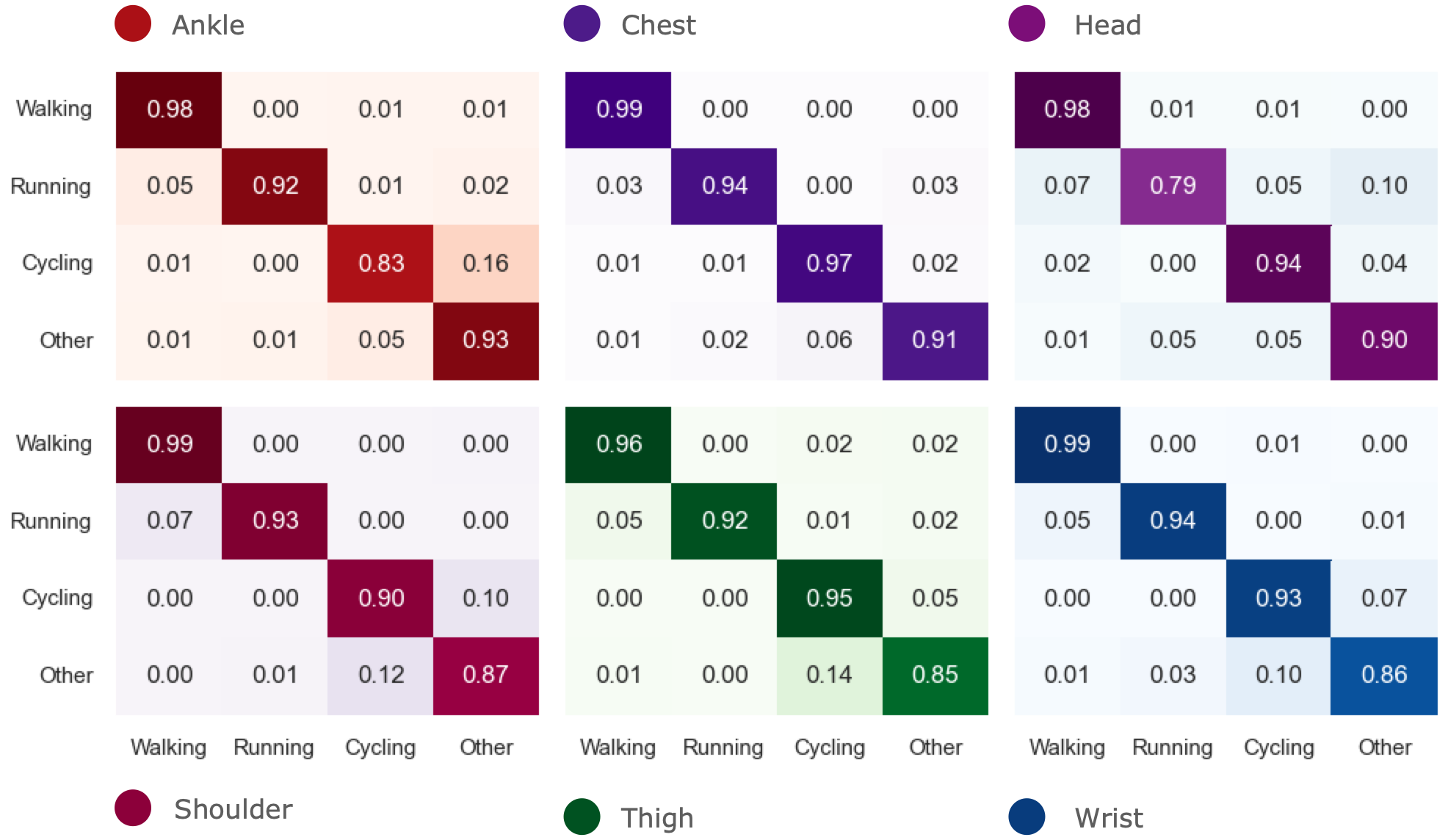}
    \caption{Detailed confusion matrices depicting the performance of our motion model, which is trained on the superset of all sensor locations, when tested on each location.}
    \label{fig:confusion_matrix}

\end{figure}

\subsubsection{Generalizability Across On-Body Locations}
We performed a comprehensive evaluation across all possible combinations of on-body locations, to better understand how a model trained on one location can "transfer" knowledge into another location. This allowed us to gain insight into how different locations are correlated, and in turn, study the limits of our framework's transfer capabilities across different locations. More specifically, we sought to answer the following questions:

\begin{itemize}
    \item How do different sensor locations correlate, and how well does a model trained on one location transfer to another location?
    \item Is there a set of key sensor locations, which we refer to as \textit{EigenLocations}, that provide enough information to create a \textit{location-invariant} motion model, comparable to using data from all six locations? 
    \item How well does our motion model perform on data captured from unseen locations?
\end{itemize}

\textit{Location Correlation.} We conducted a cross-location analysis where we investigated the performance of a model trained on data from one or more locations on another unseen location. We compared the transferabilities of the single-location models against a model trained on spectrogram images from all six locations (see Figure \ref{fig:cross_location}).

When examining single-location models, it became evident the challenges faced when dealing with changes in sensor placements. Specifically, the performance of models trained on a particular location noticeably declined when tested on different locations. Delving into some of the results, we observed that Ankle does not transfer well to other locations. Moreover, locations closer to each other, such as wrist, shoulder, and chest, showed higher correlation because of their shared biomechanics, and, as such, exhibit higher transferabilities.

By combining all sensor locations, we demonstrated a model that learns a \textit{universal representation space}, enabling the classification of the motion activities irrespective of sensor location. This model achieved an average frame-level F1-score of 91.41\% across all locations. Figure \ref{fig:confusion_matrix} visualizes the confusion matrices when testing the universal model on each sensor location separately. As seen in the confusion matrix, there is minimal confusion among the four activity classes across all sensor locations.  \\

\begin{table*}[!t]
\caption{Generalizability evaluation of a comprehensive on-body sensor location combinations.}
\label{table:cross_locations_ALL}
\centering
\small 

\begin{tabular}{@{}ccccccccc@{}}
\toprule
\myrowcolour
&\backslashbox{\textbf{Train}}{\textbf{Evaluate}} & \textbf{Wrist} & \textbf{Ankle} & \textbf{Thigh} & \textbf{Head} & \textbf{Chest} & \textbf{Shoulder} & \textbf{Average} \\
\cmidrule{1-9}
\rowcolor{maroon!20}
\textbf{Superset} & \highest{All Locations} & 90.03	&94.82&	88.37	&91.17&	91.69&	92.39	&\highest{91.41} \\
\midrule

& Wrist & 85.78	&23.28	&34.85	&75.12	&75.71	&79.63	&62.40\\
\rowcolor{maroon!5}

& Shoulder & 67.1	&22.23	&37.68	&86.83	&85.6	&91.66	&65.18 \\

& Chest  & 65.82	&32.03	&39.57&	87.43	&91.17	&85.82&	66.97\\
\rowcolor{maroon!5}

& Head & 65.92	&35.94	&50.52	&89.60	&89.42	&87.32	&69.79\\

& Ankle & 38.26	&96.00&	50.95&	35.21	&34.50&	35.40	&48.39 \\
\multirow{-6}{*}{\rotatebox[origin=c]{90}{\makecell{\textbf{Single} \\ \textbf{Locations}}}}

& Thigh  & 66.51	&78.94	&90.04	&86.68&	77.78	&71.62&	78.60\\

\midrule

& Wrist - Shoulder & 86.88	&18.07	&39.49	&84.28	&84.19	&91.70	&67.44\\
\rowcolor{maroon!5}

& Wrist - Chest & 86.61	&18.97	&36.50	&84.71&	89.94	&88.61	&67.56 \\

& Wrist - Head  & 87.17	&26.91	&38.17	&88.52	&84.67	&86.42	&68.64\\
\rowcolor{maroon!5}

& Wrist - Ankle & 86.76	&94.36	&52.60	&67.74	&72.71	&76.46	&75.11 \\

& Wrist - Thigh  & 88.33	&73.38	&89.04&	78.48	&79.88	&80.73	&81.64\\
\rowcolor{maroon!5}

& Ankle - Thigh  & 68.35	&97.00	&88.91	&79.62	&78.27	&74.04	&81.03\\

& Ankle - Head & 59.82	&96.13	&63.52	&90.37	&78.51&	82.84	&78.53 \\
\rowcolor{maroon!5}

& Ankle - Chest & 77.16	&96.07	&58.94	&87.56	&90.82	&85.14	&82.62 \\

& Ankle - Shoulder  & 83.98	&95.39	&59.60	&86.88	&89.22&	92.71&	84.63\\
\rowcolor{maroon!5}

& Thigh - Head & 67.79	&75.17&	90.22&	89.59	&82.41	&83.48	&81.44 \\

& Thigh - Chest  & 70.09	&79.74	&89.43	&87.48	&92.02	&83.89	&83.78\\
\rowcolor{maroon!10}
& \highest{Thigh - Shoulder}  & 74.01	&75.61	&90.75&	88.86&	89.46	&92.74&	\highest{85.24}\\

& Head - Chest & 67.59	&38.47	&42.16	&90.81	&92.08&	86.84	&69.66\\
\rowcolor{maroon!5}

& Head - Shoulder & 70.85	&26.80	&42.70	&90.93	&89.69	&92.73	&68.95\\

\multirow{-15}{*}{\rotatebox[origin=c]{90}{\makecell{\textbf{Double} \\ \textbf{Locations}}}}
& Chest - Shoulder  & 69.89	&29.20	&38.72&	87.85&	88.04	&89.58	&67.21 \\
\rowcolor{maroon!5}
\midrule


& Thigh - Chest - Shoulder & 73.29	&76.93	&90.47	&89.61	&92.36	&92.07	&85.79\\

& Thigh - Head - Shoulder & 74.40	&77.05	&89.45	&91.77	&90.48	&92.19	&85.89 \\
\rowcolor{maroon!5}

& Thigh - Head - Chest & 71.41 & 77.88 & 90.19 & 90.37 & 91.94 & 84.79 & 84.43 \\

& Ankle - Chest - Shoulder  &  81.74	&95.58	&59.58	&88.60	&92.05	&90.94	&84.75 \\
\rowcolor{maroon!5}

& Ankle - Head - Shoulder & 78.05	&95.66	&59.81&	91.18&	90.08	&91.57	&84.39 \\

& Ankle - Head - Chest & 73.99 & 96.21 & 57.24 & 90.29 & 91.81 & 85.37 & 82.49 \\
\rowcolor{maroon!10}
& \highest{Ankle - Thigh - Shoulder} & 84.96	&96.46	&89.56	&87.22	&89.72	&92.37	&\highest{90.05}\\

& Ankle - Thigh - Head & 69.99 & 96.44 & 89.23 & 91.18 & 82.79 & 84.36 & 85.67 \\
\rowcolor{maroon!5}

& Ankle - Thigh - Chest & 79.11 & 96.39 & 88.57 & 88.62 & 92.85 & 84.01 & 88.26 \\ 

& Wrist - Thigh - Chest  & 88.86	&72.86	&88.24	&87.82&	91.55	&86.97	&86.05 \\
\rowcolor{maroon!5}

& Wrist - Thigh - Head & 89.19	&74.69	&90.00&	88.80&	86.01	&86.09	&85.80 \\

& Wrist - Thigh - Shoulder & 90.00 & 71.49 & 88.63 & 88.88 & 86.42 & 92.17 & 86.27 \\
\rowcolor{maroon!5}

& Wrist - Ankle - Chest & 88.10	&94.79	&54.90	&83.14	&90.97	&85.94	&82.97 \\

& Wrist - Ankle - Head & 89.10	&94.57	&56.59&	88.49	&85.14	&86.48	&83.40\\
\rowcolor{maroon!5}

& Wrist - Ankle - Shoulder & 89.68 & 94.32 & 60.16 & 85.42 & 85.09 & 91.73 & 84.40 \\

& Wrist - Ankle - Thigh & 88.88 & 95.73 & 88.15 & 79.05 & 82.46 & 82.37 & 86.11 \\
\rowcolor{maroon!5}

& Wrist - Head - Shoulder & 88.03 & 16.66 & 39.86 & 88.89 & 86.86 & 92.71 & 68.84 \\

& Wrist - Head - Chest & 87.35 & 27.14 & 39.11 & 88.23 & 90.27 & 91.06 & 70.53 \\
\rowcolor{maroon!5}

& Wrist - Chest - Shoulder & 86.72 & 17.06 & 37.68 & 86.90 & 88.79 & 91.24 & 68.07 \\
\multirow{-20}{*}{\rotatebox[origin=c]{90}{\makecell{\textbf{Triple} \\ \textbf{Locations}}}}

& Head - Chest - Shoulder & 70.72 & 25.67 & 42.67 & 91.61 & 91.76 & 91.74 & 69.03 \\

\bottomrule

\end{tabular}

\end{table*}

\textit{Optimal EigenLocations.} Armed with these results and observations, we sought to ask another key question: \textit{can we find a set of sensor locations that can capture sufficient information to enable a universal motion model invariant to all possible on-body locations?} Splitting the human body into three main regions: (1) Leg Region (Ankle/Thigh), (2) Arm Region (Wrist/Shoulder), and (3) Head Region (Head/Chest), we analyzed all possible location combinations and their transferabilities. This led to the identification of \textit{EigenLocations} — a subset of locations offering the highest "transfer power." Table \ref{table:cross_locations_ALL} lists the complete evalaution results for the different single-, double-, and triple-location combinations. We observe that given two-location sets, \textit{thigh and shoulder} comprise the best double-location set, achieving an average frame-level F1-score of 85.24\% across all seen and unseen test locations, a drop of around 5\% compared to the \textit{superset} motion model. Considering triple-location sets, \textit{ankle, thigh, and shoulder} comprise the strongest EigenLocations, based on how this specific combination achieves comparable performance (90.05\% frame-level F1-score) to the model trained on all sensor locations, including locations excluded from the training. Figure \ref{fig:cross_location} depicts the performance evaluation of a subset of these combinations. \\ 


\textit{Performance on Unseen Locations.} Finally, we sought to further assess the generalizability of our \textit{location-invariant superset} model by evaluating its performance on unseen sensor locations that were not part of the training data. For this experiment, we leveraged the additional data collected from three new on-body locations (Hat, Belt, and Shoe) as described in Section \ref{sec:unseen_loc_platforms}. Our results show that the \textit{superset} motion model can generalize to unseen sensor locations achieving an average frame-level F1-score of 92.14\% across the new locations (Hat: 92.34\%, Belt: 85.77\%, Shoe: 98.31\%).\\

\textit{Performance On Unseen Platforms.}
Finally, we tested the performance of our \textit{location-invariant} deployable \textit{superset} motion model, that is trained on all locations, on an off-the-shelf sensor, using data collected from an RPi Zero, as described in Section \ref{sec:unseen_loc_platforms}. Testing our pretrained \textit{superset} motion model without any fine-tuning on this dataset, we achieved a frame-level F1-score of 98.24\% on Ankle and 85.05\% on Wrist. This experiment further strengthens our hypothesis, and demonstrates the generalizability property of our model across sensor devices. A real-time demonstration of the model running on several devices including a phone, watch, buds, and off-the-shelf hardware can be viewed in our accompanying video.     






\begin{figure}[t]
    \centering
    \includegraphics[width=.9\columnwidth]{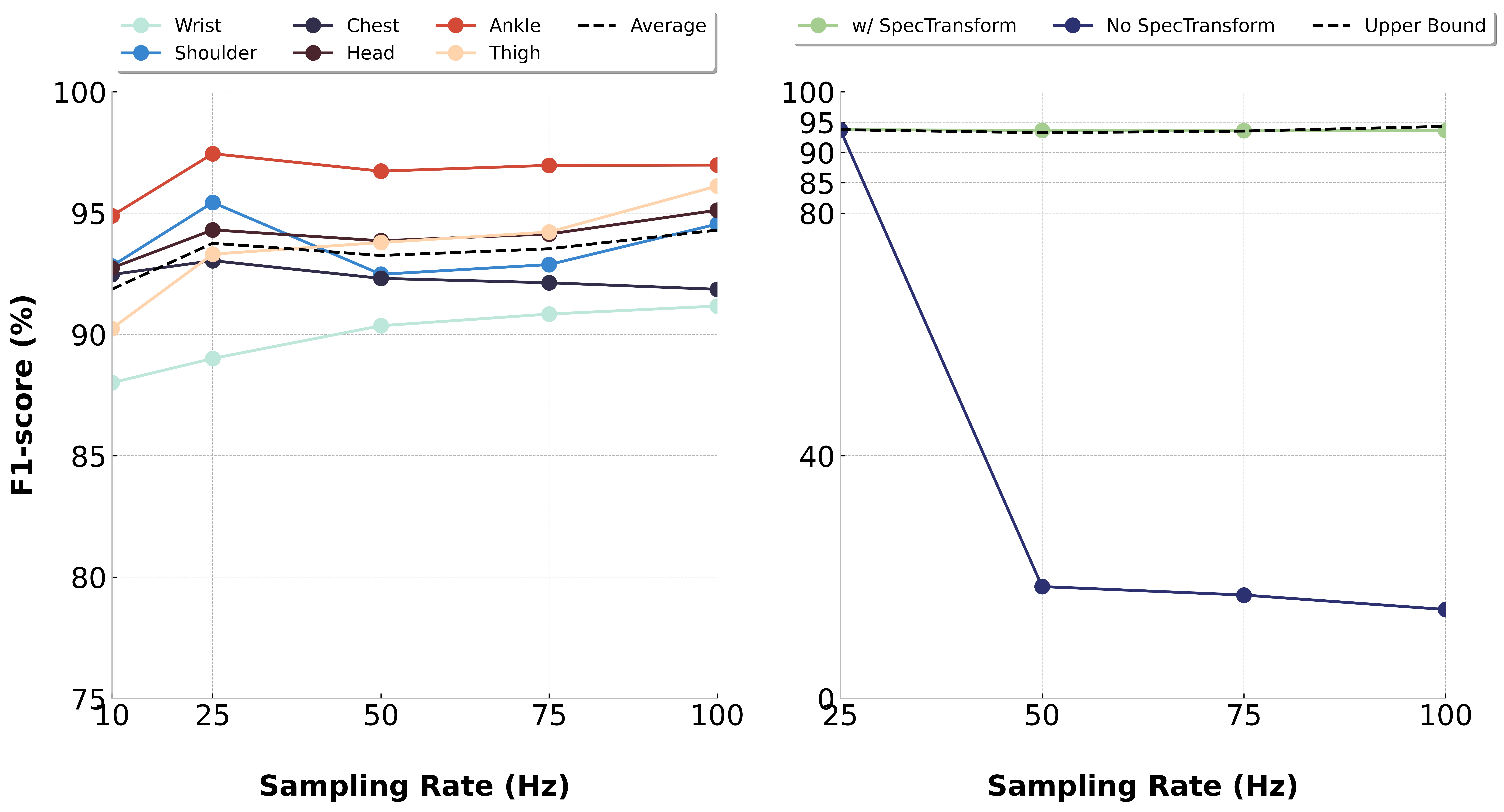}
    \caption{(Left plot) Ablation study: A universal motion model is trained on \textit{MotionPrint} data at different sampling rates and performance is reported for each location. We see that a \textit{location-invariant} motion model can be achieved at lower sampling rates. (Right plot) SpecTransform: Using a motion model trained on 25 Hz data from all locations, we show a performance curve when testing on data at different sampling rates with and without applying our SpecTransform approach. We also show that using SpecTransform achieves similar performance at each sampling rate to the corresponding universal model (upper bound).}
    \label{fig:Fs_vary}

\end{figure}

\begin{figure}[t]
    \includegraphics[width=\columnwidth]{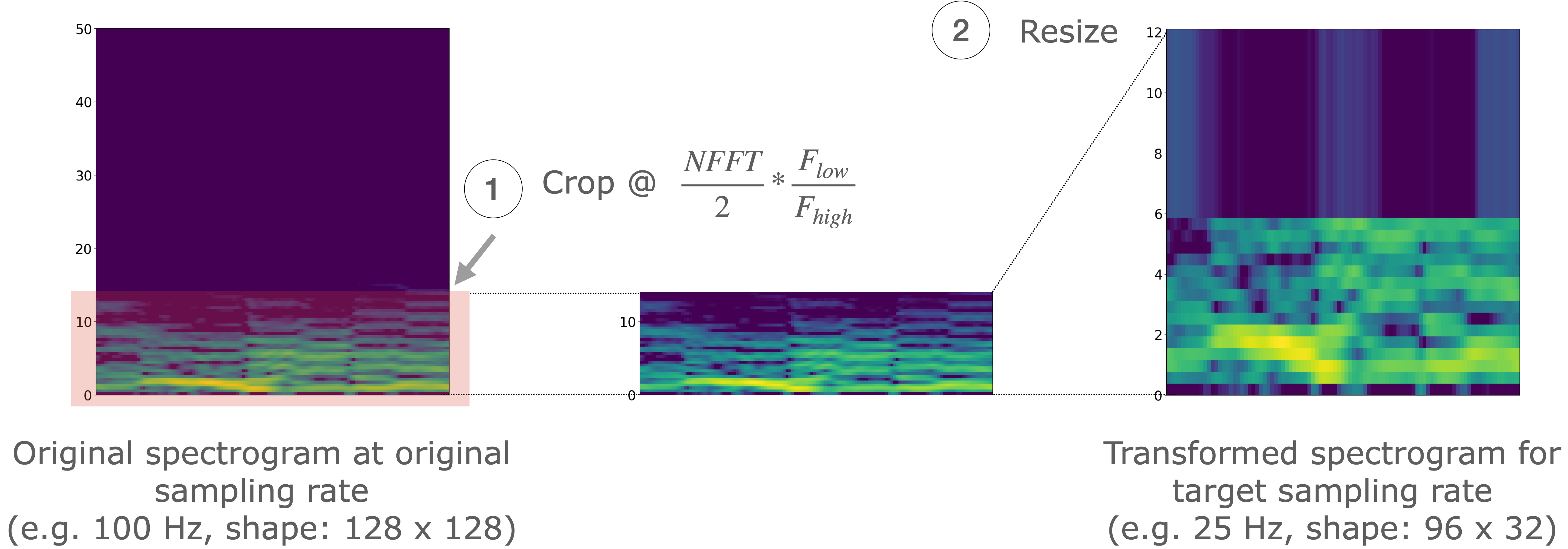}
    \caption{\textit{SpecTransform} transforms, where we apply image-based operations on spectrogram images for signal conditioning e.g., sampling rate matching.}
    \label{fig:sr_diagram}

\end{figure}

\subsubsection{Generalizability Across Varying Sampling Rates}

Sampling rate mismatch continues to be a challenge in real-world deployment of sensor-based HAR systems, where a slight mismatch in sampling rates between the training and deployment leads to a degradation in performance \cite{10.1145/3550299}. One obvious solution is to match rates by \textit{downsampling} or \textit{upsampling} in the time domain. In this analysis, we explore an alternative approach, where we perform operations in the spectrogram image domain, allowing us to borrow techniques from image processing. Specifically, we aim to answer the following questions:
\begin{itemize}
    \item Is there a performance degradation when training a location-invariant motion model with lower sampling rates?
    \item How does a sampling rate mismatch between the training data and test data affect model performance?
\end{itemize}

\textit{Ablation Study.} To answer the first question, we downsample our \textit{MotionPrint} data to different sampling rates ([100 Hz, 75 Hz, 50 Hz, 25 Hz, 10 Hz]) and train multiple motion models with all sensor locations resampled to each rate. Based on each sampling rate, we modify the hyperparameters for data segmentation and spectrogram extraction to match similar conditions to our baseline rate of 100 Hz. More specifically, the hyperparameters are transformed by multiplying the ratio $\frac{F_{new} (Hz)}{100 (Hz)}$ where $F_{new}$ represents the new target rate. Table \ref{tab:samplingRate_parameters} in Section \ref{appendix:SamplingRate} of the appendix lists the different framework parameters for every sampling rate. With this, the FFT size, and, in turn, the spectrogram input shape change. With this analysis, we show that a \textit{location-invariant} deployable motion model can be achieved at sampling rates as low as 10 Hz (Figure \ref{fig:Fs_vary}).\\

\textit{Image Operations on Spectrograms.} We also explored the effectiveness of sampling rate matching using simple image-based transformations on our spectrogram image feature sets (see Section \ref{sec:Spec_feature_set}). This expands our palette of \textit{signal conditioning} techniques by borrowing tools from computer vision and image processing. To narrow our exploration, we focus on replicating a \textit{downsampling} operation on the frequency domain, and assess the robustness of a "lowest viable sampling rate" universal model (25 Hz). For brevity, we refer to this spectrogram-based operation as \textit{SpecTransform}.

Figure \ref{fig:sr_diagram} shows the step-by-step process of SpecTransform. Revisiting the fact that a spectrogram is a visual representation of the spectrum of frequencies of a signal over time, cropping the spectrogram at a certain frequency would be conceptually equivalent to downsampling raw time series data. Using the frequency resolution (NFFT) set at the higher sampling rate (\textit{e.g.} 100 Hz), the spectrogram can be downsampled to a lower sampling rate (\textit{e.g.} 25 Hz) by cropping the spectrogram image along the frequency dimension at $c=\frac{NFFT}{2}\times\frac{F_{low}}{F_{high}}$ where $F_{high}$ and $F_{low}$ denote the current high and target low sampling rates. The spectrogram is then further resized to match the input shape of the model (\textit{e.g.} 25 Hz model has input shape $96 \times 32$).  

The effectiveness of SpecTransform was measured by simulating scenarios where data is sampled at ([100 Hz, 75 Hz, 50 Hz]) while running a model trained on 25 Hz data, the results of which can be seen in Figure \ref{fig:Fs_vary}. These results show that operations performed on an image representation (\textit{i.e.,} spectrogram) can work. In the future, we plan to investigate and borrow other image manipulation techniques, expanding the possibilities for de-noising, filtering, and augmentation.

\begin{figure*}[t]

    \includegraphics[width=\textwidth]{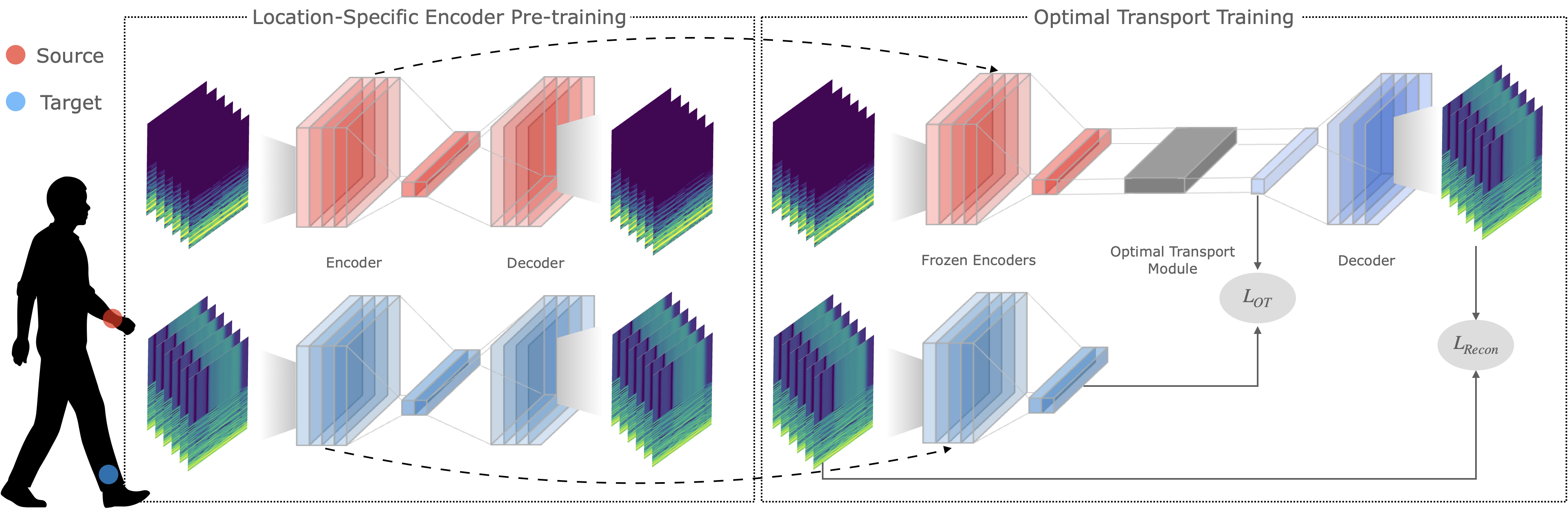}
    \caption{Our data synthesis framework. Our processs starts by first training location-specific AutoEncoders for each of the source and target locations. Next, using these pretrained encoders, an optimal transport mapping is learned using the OT Module to align the source location embedding to the target location embedding, which in turn is passed through a decoder for synthesis. During inference, the target encoder is thrown away. }
    \label{fig:OT_framework}

\end{figure*}

\section{Data Synthesis}
 In this section, we shift our focus and explore the feasibility of \textit{cross-location data synthesis}— generating synthetic motion data from one location given data from another location. As mentioned earlier, the task of collecting motion data is laborious, especially when dealing with multiple sensor placements. This technique can relieve this bottleneck.  Here, we propose mechanisms for synthetic data generation, specifically by taking into account the distribution differences observed across sensor locations.   

\subsection{Framework}
Given a source and target location, we break-up the problem of cross-location data synthesis into two steps. The first step entails training location-specific \textit{AutoEncoders} to learn a lower-dimensional representation of the sensor data for each location. Then, the pretrained encoders are frozen and employed in a Siamese-like architecture with a feature alignment module followed by a decoder for synthesis. More specifically, we exploit the \textit{optimal transport theory} \cite{cuturi2013sinkhorn} and construct a Transport Module that learns a feature transport matrix. This matrix is then applied to the embedding from the source location to transport it into the target location domain before passing it to the decoder. The decoder then generates synthetic data that would essentially correspond to the target location. The key idea is that, once this is done, given data from the source location, we can generate data corresponding to the target location.  Figure \ref{fig:OT_framework} illustrates an overview of our approach.

\subsubsection{Cross-Location Optimal Transport}
Following the optimal feature transport method previously used in cross-view image geo-localization \cite{shi2020optimal}, we leverage the same idea to learn a cross-location transport matrix that applies a feature mapping from one sensor location to another. This bridges the distribution gap between the two feature vectors, and, in turn, facilitates data synthesis. 

This module consists of computing a cost matrix between the feature vectors of the source and target locations which is then used to optimize the following objective function: 
\begin{equation}
    \textbf{P}^* = \argmin_{\textbf{P} \in P} \langle \textbf{P}\;,\;\textbf{C} \rangle_F - \lambda h(\textbf{P}),
\label{eq:OT}
\end{equation}

\noindent where $\textbf{P}^*$ is the optimal feature transport matrix we aim to learn. Equation \ref{eq:OT} is efficiently solved using the Sinkhorn-Knopp solver. 

The transportation module now aligns the feature vector corresponding to the source location to the corresponding feature vector for the target location. The transported feature is then passed to a decoder to synthesize data for the target location. We apply a mean-squared error reconstruction loss ($L_{Recon}$) between the decoder output and target location input as well as a mean squared error loss ($L_{OT}$) between the target feature vector and transported source feature vector. The total loss is expressed as:
\begin{equation}
    L = L_{OT} + L_{Recon}
\end{equation}

\begin{figure}[t]
    \centering
    \includegraphics[width=.8\columnwidth]{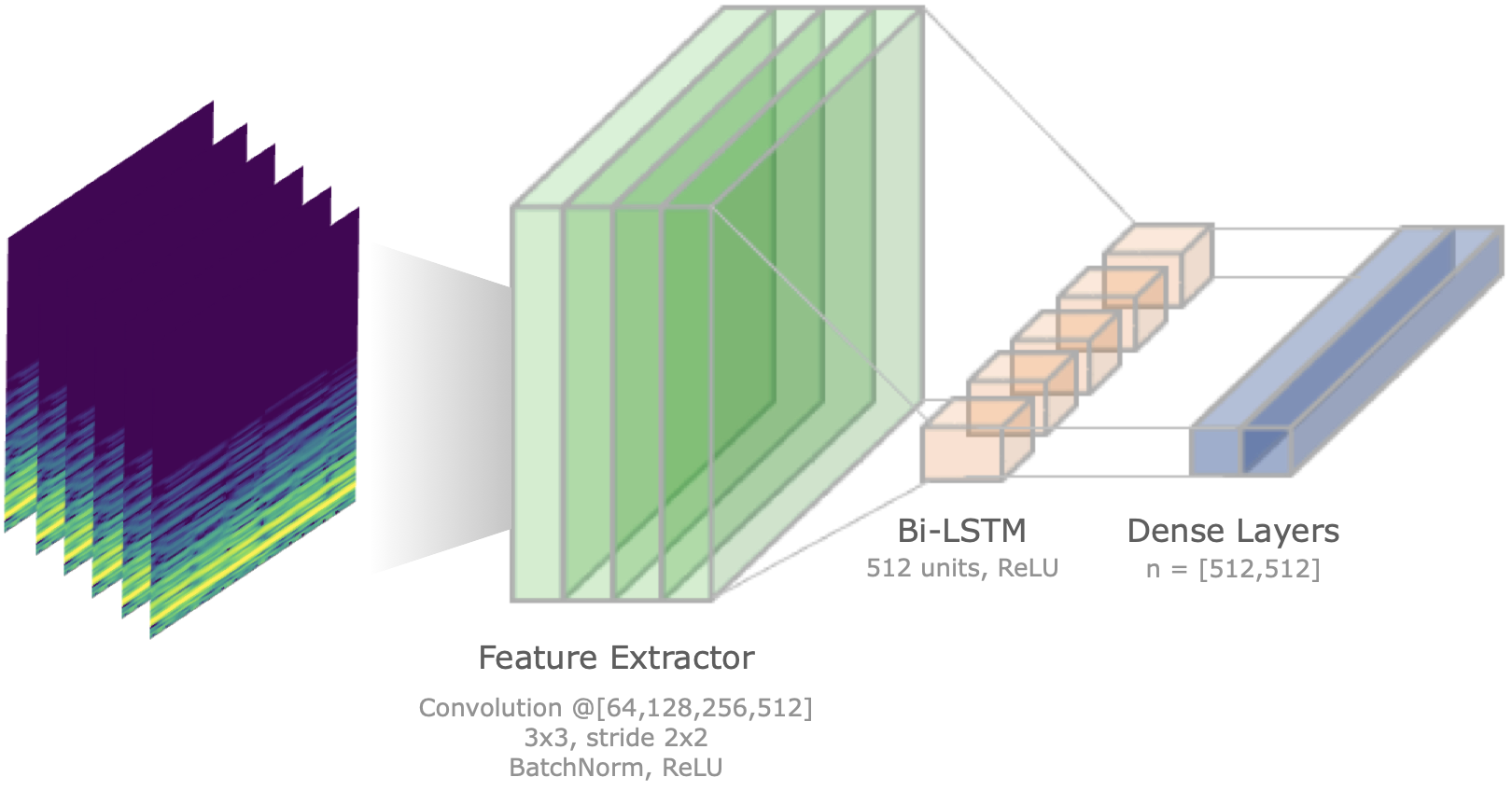}
    \caption{Encoder architecture used in Cross-Location Data Synthesis. It maps the spectrogram images to a 512-dimensional feature vector.}
    \label{fig:encoder_architecture}

\end{figure}

\subsubsection{Network Architecture}

Leveraging our spectrogram image feature sets, we again take advantage of the powerful feature representation capabilities of CNNs and long-short-term memory, and built a CNN-LSTM network for feature extraction (Figure \ref{fig:encoder_architecture}). Given a spectrogram image from a source location, the encoder extracts a 512-dimensional feature vector for each location. The transport module learns an optimal mapping of a feature vector from the source location to the target location. This transformed vector in turn facilitates the generation of the target spectrogram image via a decoder. The decoder mirrors the encoder architecture and reconstructs the target spectrogram image using the transformed source feature vector from the latent space. 

\subsubsection{Training Implementation}
Using our \textit{MotionPrint} dataset, we used the same train-test split used in Section \ref{sec:universal_train_info} and performed a train-evaluation process for each location pair. We trained our model to minimize the defined loss in Equation \ref{eq:OT} for 200 epochs using the Adam optimizer with a $1e-4$ learning rate. The model with the best results on the validation set is saved and evaluated on the testing set. 

\subsection{Evaluation}
To evaluate the feasibility of this technique, we examine our cross-location data synthesis approach on several source-target location pairs (\textit{e.g.}, wrist to ankle, head to ankle, and so on) and assess viability. Since we have six locations in our dataset (leading to 30 combinatorial pairs), we reduce our set of experiments to 5 key location pairs with the most practical utility. Our evaluation included both a visual assessment and a classification assessment using the motion model trained on the corresponding target location, essentially demonstrating whether the main activity-related information in the synthetic spectrograms is sufficiently captured.

\begin{figure*}[t]

    \includegraphics[width=\textwidth]{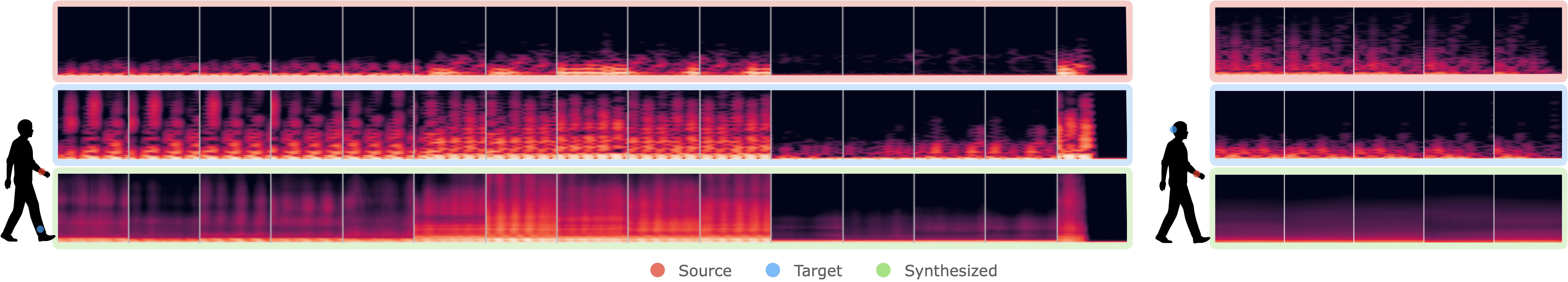}
    \caption{
    Examples of cross-location spectrogram data synthesis, showing the input spectrogram from source data (top red row), the ground truth target spectrogram (middle blue row), and the synthesized target spectrogram that was generated (bottom green row). The example on the left shows a source location of Wrist and target location of Ankle. The example on the right shows a source location of Wrist and target location of Head, which serves as a representative case where the framework can sometimes struggle to generate viable spectrograms.}
    \label{fig:spec_viz}

\end{figure*}




\subsubsection{Visual Assessment}
First, an obvious test of the effectiveness of our approach is a visual inspection of the generated spectrograms compared to the target spectrograms. For example, using our data synthesis framework trained to synthesize ankle spectrogram data from wrist data, we visualize, side-by-side, the input wrist spectrogram, the original ankle spectrogram, and the generated ankle spectrogram in Figure \ref{fig:spec_viz}. Visually, there are differences between the wrist and ankle spectrograms, and the model was able to generate the ankle spectrograms given the wrist as input. We additionally demonstrate cases of failure wherein our model is unable to generate viable spectrograms. Figure \ref{fig:spec_viz} shows an example case when synthesizing head data from wrist data. 



\begin{figure}[t]

    \includegraphics[width=.9\columnwidth]{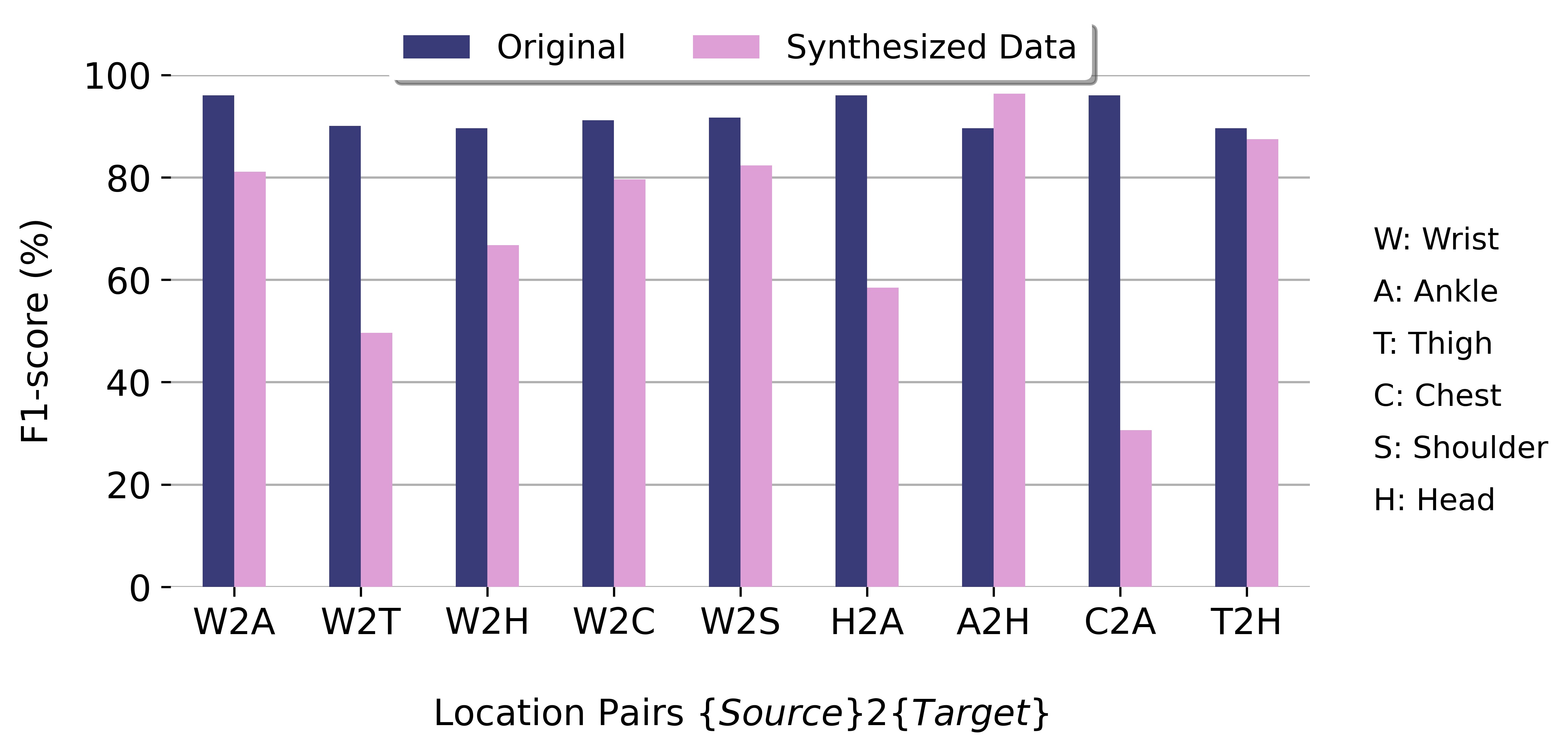}
    \caption{Performance of our motion models on synthesized spectrograms per target location. For each source-target location pair, performance on synthetic data is compared against performance on original (real) data when tested using motion models trained on each corresponding target location.}
    \label{fig:synthesis_classification}

\end{figure}

\subsubsection{Classification Assessment}
To support the utility of our cross-location data synthesis approach, we go back to our motion models and evaluate the quality of the generated spectrograms in capturing the key activity-related information. With that in mind, we test the motion model corresponding to every target location on the generated synthetic data and obtain a classification performance measure (F1-score), which demonstrates whether the model is able to recognize the activity from the generated target spectrograms. We perform this analysis on our defined location pairs and summarize results in Figure \ref{fig:synthesis_classification}. Results show that for some location pairs (Wrist2Ankle, Wrist2Chest, Wrist2Shoulder, Ankle2Head, Thigh2Head) the generated spectrograms capture the relevant activity information in the time-frequency domain resulting in more than 82\% F1-score. For other location pairs, the synthesis proved to be more challenging, resulting in non-viable spectrograms. While our proposed synthesizer falls short for some location pairs, we believe this analysis provides initial benchmarks to drive new research and accelerate advances in the field.

\section{Discussion}
\label{sec:discussion}
Here, we further extend our investigation of the potential generalizability of our motion models, discussing limitations and providing additional insights into this problem space.

\subsection{Insights from Our Generalizability Analysis Across Sensor Locations}
With our comprehensive cross-location analysis, our investigation into the different combinations of on-body sensor placements revealed notable insights. 

Analyzing the single-location models highlights the potential for achieving enhanced model generalizability through the combination of multiple sensor locations. Specifically, we observe the challenges of a model trained on one location to work \textit{off-the-shelf} on other locations. For example, a model trained on wrist data exhibits around 62\% drop in performance when tested on Ankle data compared to Wrist data. Furthermore, we observe a higher degree of correlation among locations in close proximity due to their shared biomechanical characteristics, as exemplified by a model trained on the Head achieving comparable performance when tested on Chest and Shoulder, yet experiencing a more significant decrease in performance when tested on Wrist, Ankle, and Thigh. Another intriguing finding is that constructing a model exclusively based on thigh data results in the most favorable tradeoff in performance across all locations, with an average F1-score of 78.6\%. Specifically, the most substantial decline in performance is observed when testing the model on wrist and shoulder data.

When exploring double-location combinations, we identified the \textit{Thigh-Shoulder} pairing as a promising combination, achieving 85\% average F1-score across all locations. The \textit{Ankle-Shoulder} pairing presented as a viable alternative; however, this configuration entailed a tradeoff in performance with respect to the \textit{Thigh-Shoulder} combination. More specifically, the former pairing demonstrated a challenge in generalizing to Thigh data. The latter choice, on the other hand, offered advantages through the Thigh's ability to capture a joint representation of both lower body and torso, thus achieving a better performance tradeoff between the two. Notably, configurations involving Thigh consistently achieved an average performance exceeding 80\%, underscoring the utility of including Thigh data captured by a phone in training motion models generalizable across various on-body locations. These findings provide valuable insights into optimizing sensor combinations for location-invariant motion models.

In the context of triple-location combinations, we observed that the combination of \textit{Ankle, Thigh, and Shoulder} emerged as our \textit{EigenLocations}, offering the best generalization power across all on-body locations (90.5\% F1-score). Looking closely into the other combinations, we observe that the set of \textit{Ankle, Thigh, and Chest} locations provide the next best combination with an average F1-score of 88.26\%. Including Shoulder data in our \textit{EigenLocations} instead of Chest data seems to have added additional information regarding arm movement that is not captured with the Chest data. The exclusion of Ankle from the triple-location combination results in a significant performance drop of approximately 20\%, emphasizing its importance in capturing lower-body patterns. Furthermore, in comparison to Ankle data, incorporating Thigh data in the triple-location combinations demonstrated a greater capacity to capture information conducive to the development of location-invariant models, as it encompasses joint patterns of both the lower body and torso.

Considering these collective observations, it becomes apparent that Thigh data stands out as a representative on-body sensor location that effectively captures motion information encompassing both lower body and upper body movements. This characteristic makes it particularly instrumental to the development of a motion model capable of generalization across all sensor locations. This further reinforces the practicality and relevance of collecting data using phones placed in pockets, a setting that has been widely adopted in the field.


\subsection{Insights from Our Data Synthesis}
Our evaluation of data synthesis, incorporating both visual and classification analyses, yielded valuable insights into the feasibility and challenges associated with various source-location pairs. As depicted in Figure \ref{fig:synthesis_classification}, we observed that the Wrist sensor, being the most prevalent sensor location, can effectively serve as a source for synthesizing data for the Ankle, Chest, and Shoulder, with a slight decline in F1-score in the range of approximately 10\% to 15\% when compared to the original data. An interesting observation is the 7\% improvement in performance when synthesizing Head data using Ankle data. Similarly, Thigh-to-Head data synthesis demonstrated promising results, achieving classification performance comparable to that of the original data.

In some cases, the process of data synthesis was unable to generate viable spectrograms, leading to degradation in the classification performance. An example is the generation of Head and Thigh data from wrist data, resulting in a significant reduction in classification performance, approximately 23\% and 40\%, respectively. Visually inspecting the spectrograms generated during Wrist-to-Head data synthesis (as depicted in Figure \ref{fig:spec_viz}), we observed that the generated spectrograms failed to capture the motion patterns typically observed at the target locations. Moreover, when attempting to synthesize thigh data from wrist data, we observed a higher level of noise in the thigh data due to the loose placement of the phone in a pocket.



\begin{figure}[t]

    \centering
    \includegraphics[width=.7\columnwidth]{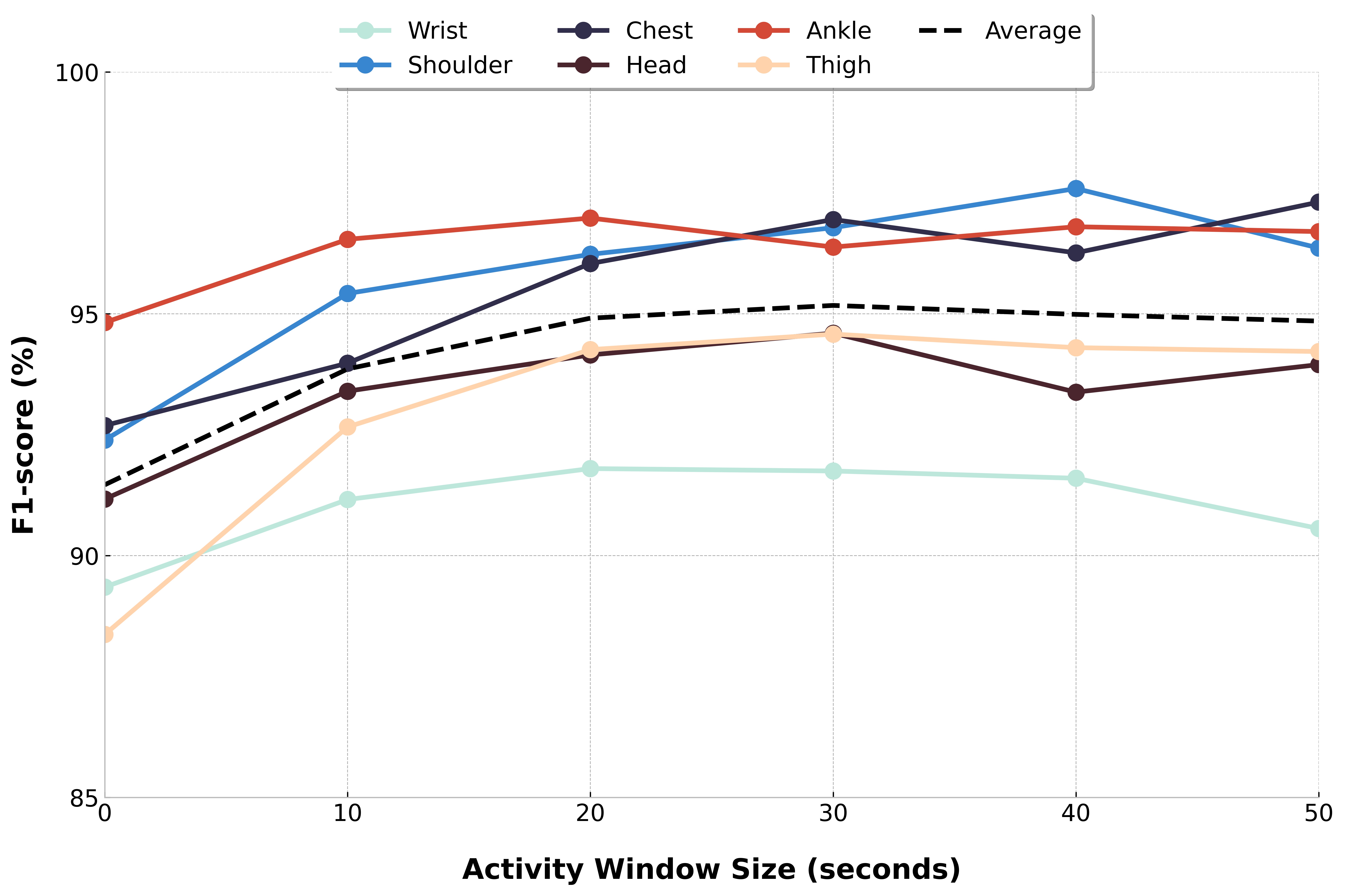}
    \caption{Performance curve of activity-level prediction of our motion model trained all sensor locations, conditioned on activity window size.}
    \label{fig:activity_level}

\end{figure}

\subsection{Activity-Level Prediction}

So far, our analysis focused on frame-level predictions, where a prediction is given for every 5.12 second frame. Practically, in real-time scenarios, \textit{sustained motion activities}, such a walking, running, or cycling, are performed longer and therefore across multiple prediction frames. Therefore, we also investigated how our frame-level predictions improve when aggregated at the activity-level. This aggregation involves deciding how many consecutive prediction frames are required before the model predicts an activity. We performed a grid search on the length of what we define as a sustained activity, tuning the parameter from 10 seconds to 50 seconds. Using our \textit{superset} model, aggregation significantly improves its recognition performance, reaching an average F1-score of 95.17\% across all locations, and going as high as 97\% when tested on Chest data, with a 30-second activity window. Figure \ref{fig:activity_level} shows the effect of varying the activity window size on the performance per sensor location. When testing the \textit{superset} motion model on data captured using RPi, applying activity-level recognition with 30-second activity window significantly boosts performance on Wrist data, reaching 93.02\% and maintains performance on Ankle data with 98.3\%.

\subsection{Motion Embeddings}
\label{sec:embeddings}
So far, we've demonstrated various aspects of our large-scale \textit{MotionPrint} dataset that allowed us to (1) build readily-deployable motion models that can work across devices and sensor placements, and (2) explore cross-location data synthesis. In this section, we further investigate another characteristic of our \textit{superset} motion model, specifically its ability to extract powerful generalizable motion representations that can be fine-tuned and transferred to any downstream task. We evaluate the transferability of our embeddings to other target datasets of different activity classes. By providing the community with our pretrained models and pretrained embeddings, we envision our work facilitating a wide range of applications.


To conduct this experiment, we followed the standard \textit{pretrain-finetune} paradigm of transfer learning where, given our pretrained \textit{superset} motion model from Section \ref{sec:universal_model}, we freeze the learned weights of the first-stage feature extractor and fine-tune the 3-layer MLP classifier to a target dataset. The number of neurons of the last layer of the classifier is modified according to the corresponding target downstream task. For all experiments, training of the MLP classifier was done using an Adam optimizer with a learning rate of 0.0001.   \\


\begin{figure}[t]

    \includegraphics[width=.8\columnwidth]{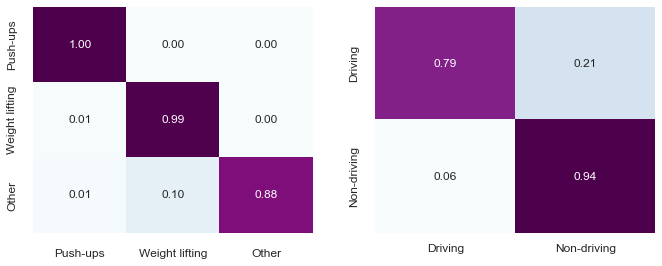}
    \caption{Confusion matrix results, after applying Transfer Learning of physical activity (left) and driving (right) classification tasks using the motion embeddings from the \textit{superset} motion model.}
    \label{fig:TL_cm}

\end{figure}

\subsubsection{Physical Activity Classification} 
To investigate the transferability of our motion embeddings to other unseen classes, we filter out two activities (push-ups and lifting weights) from our original "other" class, to serve as our target dataset for activity classification. In this particular test, we build a variant of our universal model trained on source data that excludes these two classes. In this analysis, we focus on wrist data to better simulate a practical real-world use of our embeddings. Each of the source and target datasets are split by participants into train, test, and validation sets. We also re-purposed random samples from other non-pushup and non-weightlifting activities into a new "other" class (3-class total: pushups, weighlifting, other). Overall, this embedding-based model (after fine-tuning) achieves a frame-level F1-score of 85.93\% , confirming our hypothesis that our embeddings can be extended to other downstream tasks given minimal data. Figure \ref{fig:TL_cm} (left) shows the confusion matrix from our test set. \\

\subsubsection{Driving Classification}
To further validate our motion embeddings, we also performed a similar analysis on a completely different dataset and recognition task: driving detection. Leveraging a separate driving dataset captured using an Apple Watch, we investigate the transferability of our motion embeddings towards building a driving classifier. The dataset includes driving and non-driving activities (walking, cycling, etc.), balanced roughly equally. Our results show that our embeddings-based driving classifier achieves a frame-level F1-score of 87.16\% (Figure \ref{fig:TL_cm}, right). Testing this model in real-time shows that we can use it to detect if a user is in a moving car, regardless if they are the driver or the passenger.  This is encouraging, as we show that the motion embeddings learned from our \textit{superset} \textit{location-invariant} motion model can extract features that generalize to unseen activities. A real-time demonstration of these two applications are included in our accompanying video. 

\subsection{Limitations}
While this paper provides a comprehensive exploration for advancing location-invariant and device-agnostic motion recognition models, it is important to highlight key limitations and potential avenues for improvements.

First, our set of motion classes are purposefully narrow, targeting sustained activities that are mobile in nature. Although highly impactful, future work is needed to better understand how to translate the generalization properties of our model to other activities. Studies have shown that the optimal sensor location depends on the target activity to be recognized \cite{atallah2011sensor}. For instance, unlike locomotion, some activities are only "felt" on certain body-parts, e.g. transient activities, static poses, or even gestures. Future work could delve into these areas to enhance our understanding and broaden the applicability of activity recognition models to a wider range of activities and sensor placements.

It is important to acknowledge that our \textit{MotionPrint} dataset, although extensive, does not include individuals with accessibility needs, such as those who use mobility aids like wheelchairs or canes.
Therefore, it is uncertain whether our \textit{ready-to-use} deployable motion model can be readily applied to users with such accessibility requirements. Each user's accessibility condition is unique, making it necessary to leverage our motion embeddings (Section \ref{sec:embeddings}) and fine-tuning techniques to develop a personalized location-invariant model. Exploring this critical problem domain is part of our future research plans. 

Likewise, in this work, we focused on the most ubiquitous form factors that people use daily (\textit{e.g.,} watches, phones, and headsets). Additionally, we only focused on on-body accelerometers due to their practicality, as they are omnipresent in most wearable devices and highly power-efficient. Nevertheless, modalities beyond accelerometers (\textit{e.g.} gyroscope, audio, images) in other form-factors have been proven effective in activity recognition tasks \cite{OkGoogle,laput2018ubicoustics}. Combining modalities can provide useful additional information that can improve performance \cite{10.1145/3534582} in the future. 







\subsection{Applications}
In this final section, we show how our contributions can be highly impactful in enabling various end-user applications. Please see our video figure for more details and examples.

\subsubsection{Motion-Aware Cameras}
Handheld and on-body cameras (\textit{GoPros}) can be great tools for capturing high-action activities like bike rides and runs. Even DSLR cameras are now equipped with motion sensors for optical image stabilization, offering "action" capabilities. By integrating our \textit{ready-to-use}, \textit{location-invariant} motion model into these cameras, enhanced mode adjustments can be made based on the user's motion context. For example, when used with body-mounted cameras, our model can provide automatic activity recognition to enable “Activity Mode” switching, loading custom camera profiles during bike rides or runs. 




\subsubsection{Context-Aware Devices}
Most non-desktop workstations today (\textit{e.g.,} laptops and tablets), are equipped with accelerometers. Thus, integrating our model into such devices can enable new interactive opportunities that leverage real-time context-awareness. For example, laptops can be made aware of their surroundings when being carried by the user while walking and thus adjust its settings accordingly. This could include increasing the font size or adjusting the screen brightness to make it easier to view while walking. Additionally, a similar mode can be done as a Text Message Status modifier, for example, showing that the sender is Walking while Typing. 

\subsubsection{Sports Performance Analysis}
Our large-scale multi-device multi-location dataset can unlock many research directions for fitness. For example, sports performance tracking systems are crucial in providing deep insights into an athlete's movements. This is done by capturing an athlete's full-body motion, which has been vital in the field of biomechanics, gait analysis, sports science, and even animal research. We believe our \textit{MotionPrint} dataset, cross-location analysis, \textit{location-invariant} motion model, and data synthesizer, can move the ball forward (pun intended) towards advancing research in sports performance analysis. 







\subsubsection{Mixed-Reality Motion Sensing}
Motion sensing is a fundamental component of most mixed-reality experiences. For example, on-body VR equipment such as wearable head-mounted displays or even handheld controllers such as the Playstation Move or Oculus Touch already contain several motion sensors\cite{boas2013overview}. We envision our \textit{ready-to-use} motion model seamlessly integrating into various on-body mixed-reality equipment, thus further advancing this immersive experience. Other kinds of toys equipped with accelerometers can also be upgraded with motion-activated features enabling numerous fun experiences. 

\subsubsection{Motion-based User Identification}
Leveraging our \textit{MotionPrint} dataset, another interesting application in the field of user authentication can be explored. More specifically, by monitoring motion data from several devices, \textit{e.g.} capturing the walking pattern from the earbuds and watch placed on the wrist, it is possible to identify whether both motion patterns correspond to the same person \cite{gong2022user}. A correlation analysis of the motion patterns captured simultaneously from several devices can provide an identification metric that authenticates the user. With our multi-device multi-location dataset, an in-depth exploration of this application can be conducted across several on-body locations. This can further push this notion of a \textit{digital motion footprint} of a user. 




\subsubsection{Real-time Automatic Annotation Tools}
Using our \textit{location-invariant} deployable motion model as a real-time automatic annotation tool can be useful for researchers to label accelerometer data collected from any sensor-location or device. This can help quickly and efficiently annotate large amounts of data, which would otherwise require a significant amount of manual labor, thus streamlining the data collection and annotation process. This can also be used to annotate other modalities that are collected simultaneously, such as video and audio. 




\section{Conclusion}
In this paper, we present the results of our comprehensive investigation on the generalizability of motion recognition models across diverse on-body sensor placements and devices. As a result, we identify representative sensor locations that facilitate the development of \textit{location-invariant} models that can be integrated into any device with \textit{one} accelerometer placed anywhere on the body, achieving an average frame-level F1-score of 91.41\% across various on-body locations. Aggregating at the activity-level further improves the recognition performance, reaching an average of 95.17\% F1-score using a 30-second activity window. This yields a plug-and-play system that requires no additional end-user training or calibration. In order to achieve this, we also present and make public our \textit{MotionPrint} dataset, a strongly-labeled multi-device multi-location sensor-based human activity dataset -- the largest of its kind.  Additionally, we introduce \textit{cross-location data synthesis}, where we explore and  demonstrate its promising capabilities for generating synthetic data from one on-body location to another. Finally, we make our dataset, \textit{ready-to-use} and \textit{location-invariant} models, and synthesizer recipes publicly available, to drive new research, lowering the floor for researchers and practitioners to make progress in this field. Ultimately, we hope this work advances research and creates impactful applications in ubiquitous computing and interactive systems.

\section{Acknowledgments}
We would like to thank all participants for contributing to our dataset, and all study coordinators for hosting the study. 